\documentclass[conference]{IEEEtran}
\IEEEoverridecommandlockouts
\usepackage{cite}
\usepackage{amsmath,amssymb,amsfonts}
\usepackage{algorithmic}
\usepackage{graphicx}
\usepackage{textcomp}
\usepackage{xcolor}
\def\BibTeX{{\rm B\kern-.05em{\sc i\kern-.025em b}\kern-.08em
    T\kern-.1667em\lower.7ex\hbox{E}\kern-.125emX}}

\usepackage{xr}
\usepackage{soul}
\usepackage{makecell}
\usepackage{booktabs}
\usepackage{enumitem}
\usepackage{multirow}

\ifCLASSOPTIONcompsoc
    \usepackage[caption=false,font=normalsize,labelfon
    t=sf,textfont=sf]{subfig}
\else
    \usepackage[caption=false,font=footnotesize]{subfig}
\fi

\definecolor{darkgreen}{HTML}{59B02E}
\definecolor{darkred}{HTML}{B87213}

\newcommand{\drawcomments}{}
\newcommand{\revision}[1]{{#1}}

\ifdefined\drawcomments
\newcommand{\byron}[1]{\textcolor{purple}{[Byron: #1]}}
\newcommand{\lili}[1]{\textcolor{cyan}{[Lili: #1]}}
\newcommand{\sarra}[1]{\textcolor{orange}{[Sarra: #1]}}

\newcommand{\delete}[1]{\textcolor{red}{\expandafter\st\expandafter{#1}}}
\newcommand{\newcontent}[1]{\textcolor{orange}{#1}}
\newcommand{\sub}[2]{\expandafter\delete\expandafter{#1} \newcontent{#2}}

\newcommand{\agent}{MIMIC }
\newcommand{\agentIn}[1]{{MIMIC#1}}

\newcommand{\LLMP}{MIMIC-P }
\newcommand{\LLMPIn}[1]{{MIMIC-P#1}}

\newcommand{\LLMPS}{MIMIC-P+S }
\newcommand{\LLMPSIn}[1]{{MIMIC-P+S#1}}

\else
\DeclareRobustCommand{\byron}[1]{%
\ignorespaces
}
\DeclareRobustCommand{\lili}[1]{%
\ignorespaces
}
\DeclareRobustCommand{\sarra}[1]{%
\ignorespaces
}
\DeclareRobustCommand{\delete}[1]{%
\ignorespaces
}
\DeclareRobustCommand{\safedelete}[1]{%
\ignorespaces
}
\DeclareRobustCommand{\safedelete}[1]{%
\ignorespaces
}
\DeclareRobustCommand{\newcontent}[1]{#1}

\fi

\begin{document}

\title{MIMIC: Integrating Diverse Personality Traits for Better Game Testing Using Large Language Model}

\author{\IEEEauthorblockN{Anonymous Author(s)}}

\author{\IEEEauthorblockN{Yifei Chen}
\IEEEauthorblockA{\textit{Electrical and Computer Engineering} \\
\textit{McGill University}\\
Montréal, Canada \\
yifei.chen@mail.mcgill.ca}
\and
\IEEEauthorblockN{Sarra Habchi}
\IEEEauthorblockA{
\textit{Cohere}\\
Canada \\
sarra.habchi@cohere.com}
\and
\IEEEauthorblockN{Lili Wei$^1$}
\IEEEauthorblockA{\textit{Electrical and Computer Engineering} \\
\textit{McGill University}\\
Montréal, Canada \\
lili.wei@mcgill.ca}
}

\maketitle
\footnotetext[1]{Lili Wei is the corresponding author.}

\begin{abstract}

    Modern video games pose significant challenges for traditional automated testing algorithms, yet intensive testing is crucial to ensure game quality. To address these challenges, researchers designed gaming agents using Reinforcement Learning, Imitation Learning, or Large Language Models. However, these agents often neglect the diverse strategies employed by human players due to their different personalities, resulting in repetitive solutions in similar situations. Without mimicking varied gaming strategies, these agents struggle to trigger diverse in-game interactions or uncover edge cases.

    In this paper, we present \agentIn{,} a novel framework that integrates diverse personality traits into gaming agents, enabling them to adopt different gaming strategies for similar situations. By mimicking different playstyles, \agent can achieve higher test coverage and richer in-game interactions across different games. It also outperforms state-of-the-art agents in Minecraft by achieving a higher task completion rate and providing more diverse solutions. These results highlight \agentIn{'s} significant potential for effective game testing.
    
\end{abstract}

\begin{IEEEkeywords}
Artificial Intelligence, Human-Like Gaming Agents, Personality-Driven Gaming Agents, Automated Game Testing, Large Language Models (LLMs).
\end{IEEEkeywords}

\section{Introduction}\label{sec:introduction}
Modern video games have become one of the most significant entertainment sectors,  generating USD 183.9 billion globally in 2023~\cite{GameIndustryStatistics}. To maintain game quality, rigorous testing has become essential, reflected in the growth of the game testing services market, valued at USD 321.4 million in 2025 and projected to reach 670.5 million by 2033~\cite{GameTestingStats}.

Yet modern games pose significant challenges for traditional automated testing~\cite{SurveyVideoGameTesting}. A common technique, “record and replay”, where human interactions are captured and reused as test cases~\cite{RecprdAndReplay}. While effective at reproducing known scenarios, this approach often fails in nondeterministic gaming environments or when games evolve~\cite{AgentBasedAutoGameTesting}. As a result, frequent updates to the recordings are required, making it inefficient for modern game development cycles.

To address this limitation, researchers have explored agent-based testing leveraging machine learning (ML) techniques such as reinforcement learning (RL)~\cite{RLTesting1} and imitation learning (IL)~\cite{ILTesting2}. While effective at executing test plans, RL depends on rigid reward functions and IL relies on expert demonstrations, limiting its generalization to new tasks or games~\cite{61AutoMC2024}. More recently, Large Language Models (LLMs) have been applied to gaming agents. And such LLM-based agents have demonstrated impressive adaptability in solving complex tasks across diverse games~\cite{OpenAI5DefeatsDota2Team, Voyager2023, GITM2023, ChessGPT2023, startCraftRL2019}.

A common limitation of both ML- and LLM-based agents is overlooking the diverse strategies players adopt for the same task, shaped by their personalities. Humans may approach a task conservatively or aggressively~\cite{AggressivePlayerBehaviour2008}, but existing agents often ignore such behavioural diversity and generate repetitive solutions. This limits their ability to thoroughly explore games or uncover edge cases, reducing effectiveness in game testing.

To address this challenge, we propose \agentIn{,} an LLM-based framework that mimics different gameplay personalities to generate diverse solutions for the same in-game tasks and achieve higher coverage. Our key insight comes from real-world gameplay, where players may approach the same tasks with varied strategies shaped by their personalities. Yee et al.~\cite{WCPersonalityBehaviour2011} found significant correlations between personality traits and behaviours in \textit{World of Warcraft}. Similarly, Narnia et al.~\cite{WCPersonalityBehaviour2014} showed that in-game player behaviours align with real-world personality traits. For example, emotional players may prefer levelling alone to reduce negative feedback from others, whereas extroverts prefer social questing. Inspired by these observations, \agent integrates diverse personality traits into gaming agents to simulate realistic, diverse behaviours.

\agent leverages LLMs to align agent behaviour with specific personality traits. For example, when facing an opponent, a cautious agent may avoid combat, while an aggressive one would attack directly. A Memory System further records past gameplay and retrieves useful experiences, allowing agents to accumulate knowledge over time and consistently solve complex tasks in line with their personalities.

To assess \agentIn{'s} effectiveness, we use it to test two open-source games of varying complexity, measuring both \textit{code-level} and \textit{interaction-level} coverage. In the small-scale game, \agent showed performance comparable to human testers, reaching \textbf{100\% \textit{combinatorial coverage}}, which measures how thoroughly agents explore combinations of in-game actions and parameters, and narrowing the gap in code-level coverage. In the large-scale game, it consistently outperformed random-based baselines, achieving up to \textbf{1.30$\times$} higher branch coverage and \textbf{14.46$\times$} greater combinatorial coverage.

We further evaluated \agent in the widely used real-world game \textbf{Minecraft}~\cite{Minecraft}, comparing it against the state-of-the-art agent \textbf{ODYSSEY}~\cite{56Odyssey2024}. Both were assigned the same suite of in-game tasks. \agent not only outperformed ODYSSEY in task completion but also showed greater behavioural diversity in six of eight tasks, yielding broader coverage of gameplay scenarios. The key contributions of this paper are:

\begin{itemize}[leftmargin=1.4em, itemsep=2pt, topsep=0pt]
    \item We propose a novel framework that integrates gaming agents with diverse personalities, enabling more diverse and effective game testing.
    \item We conduct two studies across three games and demonstrate \agentIn{'s} effectiveness in solving complex tasks and achieving higher coverage through diverse solutions.
    \item To facilitate future research, we made \agent and all the used prompts public~\cite{MIMIC_Webpage}.
\end{itemize}

\section{Background} \label{sec:background}

\subsection[RAG for LLMs]{Retrieval-Augmented Generation for LLMs}\label{subsec:background-rag-for-llms}

Large Language Models (LLMs) are deep learning systems with billions of parameters trained on massive datasets. They generate human-like text and code with contextual and semantic awareness. Public interest surged after the release of ChatGPT, which reached about 180 million users in 2024~\cite{LLMdata2024}.

However, LLMs are prone to ``hallucinations''~\cite{marcus2020decadeaistepsrobust}, plausible but incorrect or fabricated content that may deviate from user inputs~\cite{huang2023survey}.
To mitigate hallucinations and improve efficiency, Retrieval-Augmented Generation (RAG) was introduced~\cite{11RAGOriginalPaper}. RAG has two steps: \textit{Retrieval}, which queries a predefined database for relevant information, and \textit{Generation}, which combines the retrieved content with the user query as input to the LLM. This approach improves accuracy in question answering, and has become widely studied for its efficiency and flexibility~\cite{YuHaoRAGEvaluation2024, RAGSurveyGAo2024}.

\subsection{Modelling Gamer Traits }\label{subsec:background-gamer-personality-definition}

In 1996, Bartle categorized player behaviours into four personality traits~\cite{bartle1996hearts}, shaping how players engage with games. Subsequent research expanded this work, modelling gamer behaviours through personality-based classifications~\cite{69KAHN2015354, 70TowardATheoryMotivatingInstructionMTW, 71NACKE201455, 72DefiningPersonasInGame, 73GamePreference2017, 74StrategyPrediction2009}. A recent study synthesized these efforts into seven traits: \textit{Achievement}, \textit{Adrenaline}, \textit{Aggression}, \textit{Caution}, \textit{Completion}, \textit{Curiosity}, and \textit{Efficiency}~\cite{ArtificialPlayers2020}. We leverage these traits to prompt our agents to mimic human behaviours.

\section{Motivating Example} \label{sec:motivation}

In this section, we use a task from Minecraft as a motivating example to discuss the limitations of existing work and highlight the motivation of \agentIn{.}

The task \textit{Obtain 1 diamond} is long-hailed as a significant challenge in the community~\cite{81DEPS} and serves as the focus of the NeurIPS MineRL Competition~\cite{MineRL2019}. It requires completing at least 13 sequential sub-goals, each with multiple variants, making it a long-horizon task that typically takes humans over ten minutes to solve~\cite{49Jarvis-1_2023}. Players must also handle dynamic requirements posed by the game, such as hunger or safety, further expanding the solution space.

Existing gaming agents are optimized for task completion, often producing homogeneous, repetitive behaviours. For example, in our evaluation, we observed that ODYSSEY~\cite{56Odyssey2024}, a state-of-the-art LLM-based agent, consistently followed a single optimized path to obtain a diamond, regardless of environmental variations. While such agents achieve high task completion rates, their behaviours diverge from human players, who rarely pursue tasks in strictly optimized ways. Instead, human players exhibit adaptive and diverse behaviours in response to spontaneous in-game events, shaped by their personality traits~\cite{WCPersonalityBehaviour2011, WCPersonalityBehaviour2014}. For example, an aggressive player may fight creatures for rewards while pursuing the diamond task, even if it does not directly advance the primary goal. As a result, the existing gaming agents fail to emulate this diversity and adaptability, limiting their ability to cover the wide range of unpredictable in-game scenarios and reducing their overall effectiveness for game testing.




This motivates us to propose \agentIn{,} an LLM-based agent framework that integrates personality traits into the core planning process. Unlike existing agents that narrowly pursue optimal action sequences, \agent leverages recent advances in LLMs capable of simulating consistent personality traits~\cite{78LLMPotentialOnPersonality2024, 79PersonalityTraitsLLM2023, 80LLMSimulateBigFivePersonality2024, 2022WhoIsGPT3} to model gameplay behaviours that more closely resemble those of real human players. By conditioning its dynamic Planner on distinct personality prompts, \agent generates strategies that are task-oriented and driven by personality-specific tendencies. This enables it to pursue goals while continuously responding to in-game events in a manner consistent with a player of that personality, leading to more diverse, realistic, and interaction-rich testing trajectories.
For example, in finishing the \textit{Obtain 1 Diamond} task, our \textit{aggressive} agent dedicated 21.52\% of its actions to combat, frequently upgrading armour and engaging a variety of creatures. 
In contrast, the \textit{cautious} agent avoided combat entirely, prioritizing safety by crafting torches before mining. Meanwhile, the \textit{adrenaline-seeking} agent actively crafted swords and explored high-risk areas to encounter enemies, reflecting a strong preference for challenge-oriented interactions.
These results demonstrate that by integrating personality into our gaming agent, \agent can generate meaningful actions in response to various environments according to the specified personality, driving exploration towards more diverse scenarios.

Furthermore, human decision-making is shaped not only by personality but also by experience and preferences~\cite{Analogy1983, SexualHarassment2024, PredictablyIrrational2008}. To model this, we introduce a Memory System that records past actions and outcomes as memories. The relevant and preferred ones are then retrieved to guide consistent, human-like decisions.

By combining personality-driven planning with memory-aware decision-making, \agent delivers a novel testing framework that mimics the behavioural diversity of human players and enables broader exploration of in-game scenarios.




\section{Approach}\label{sec:approach}

\begin{figure}[!t]
    \centering
    \includegraphics[width=\linewidth]{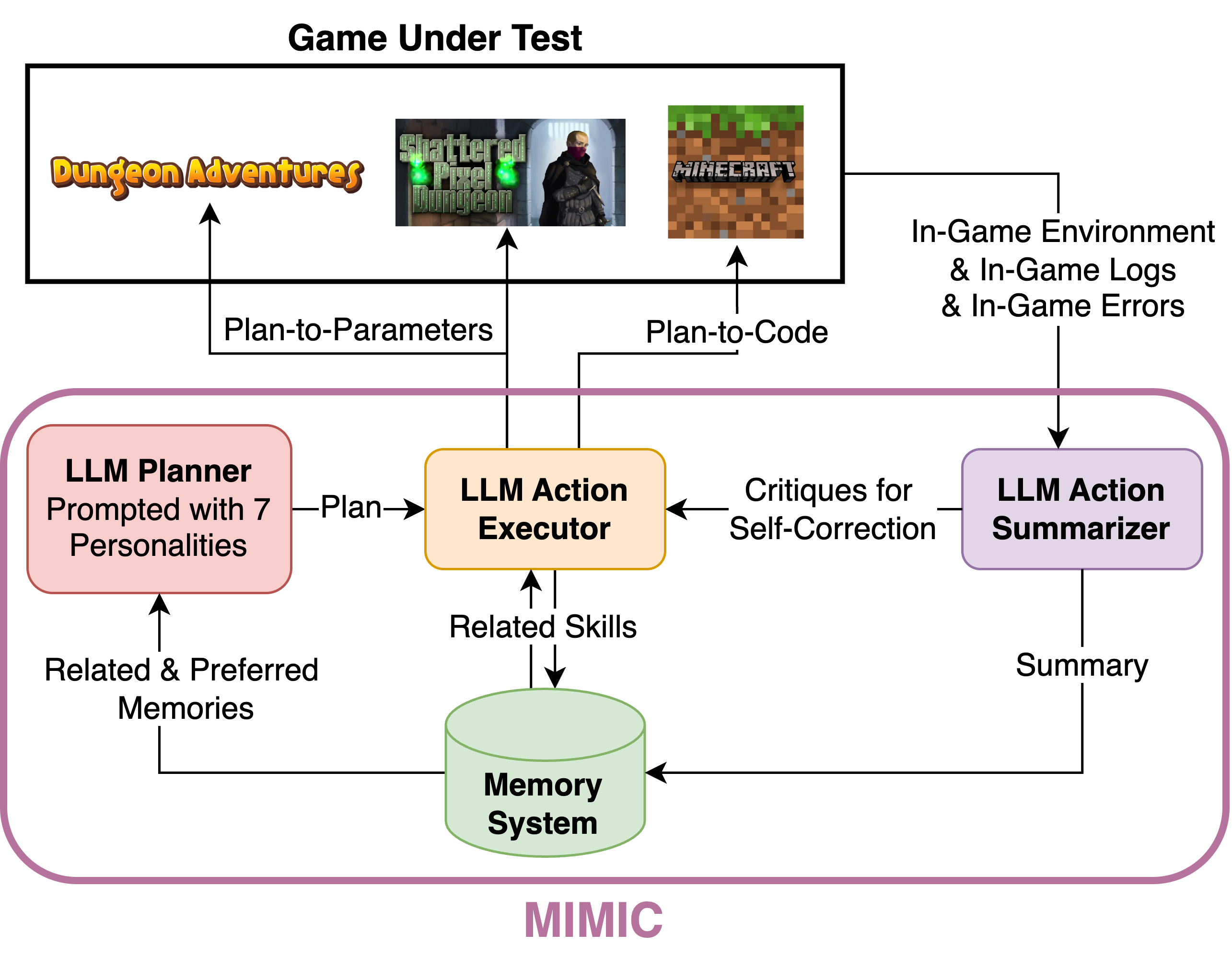}
    \caption{Overview of the \agent framework, comprising three LLM-based components: LLM Planner, LLM Action Executor, and LLM Action Summarizer, alongside a non-LLM-based component, the Memory System.}
    \label{fig:overview}
\end{figure}

\figurename~\ref{fig:overview} presents an overview of the \agent framework, which consists of four components: the Planner, Action Summarizer, Action Executor, and Memory System. 

The Planner is the core module, generating action plans from predefined personality traits and past experiences. These experiences are stored in the Memory System, with the Action Summarizer analyzing execution results to produce summaries. The Action Executor then translates the Planner’s output into in-game actions. In each action iteration, the Planner produces a plan, the Executor executes it, and the Summarizer records feedback as new memory to guide future planning. The following sections detail each component.

\subsection{Planner}\label{subsec:approach-planner-prompted-with-personalities}


The LLM Planner is the core component that integrates personality traits into the decision-making process of \agentIn{.} Unlike prior agents that follow an optimal action sequence, our planner is prompted with personality traits to generate plans that reflect diverse human-like behaviours. These plans simulate how players with different personalities may approach the same goal, enabling more varied and realistic gameplay.

To handle both immediate actions and long-term objectives in accordance with the personality traits, we adopt a hybrid planning design. This allows the Planner to dynamically alternate between low-level reactive planning and high-level goal decomposition, enhancing both adaptability and personality alignment. This hybrid design increases behavioural diversity and improves test coverage in complex game environments. 

In addition to personality traits, the Planner also receives information about the in-game environment and game mechanics, enabling it to generate grounded, context-aware plans. To support coherent progression, previously executed plans are also provided as reference (retrieved from the Memory System in Section~\ref{subsec:approach-text-based-memory-system}), allowing the Planner to build on past decisions and maintain continuity throughout gameplay.



\subsubsection{Mimicking Gamer Personality}\label{subsubsec:approach-gamer-personality-mimicking}

\revision{
    To integrate personality-driven behaviours into \agentIn{,} we find a personality model to guide it. We leveraged the model from PathOS~\cite{ArtificialPlayers2020}, which synthesizes seven personality traits, \textit{Achievement}, \textit{Adrenaline}, \textit{Aggression}, \textit{Caution}, \textit{Completion}, \textit{Curiosity}, and \textit{Efficiency}, from nine player modelling studies spanning 1981–2017. These traits are behaviourally grounded, well-defined, and generalizable, making them directly applicable to \agentIn{.}
    
    Alternative personality models are less suitable. Generic models (e.g., Big Five, MBTI) are insufficient to capture gameplay behaviours. Some game-specific models exist, but are narrower than PathOS. For example, Narnia et al.~\cite{WCPersonalityBehaviour2014} derived a model from a single game, limiting generalizability. A later study~\cite{n1} defined four personality traits, all of which are subsets of PathOS, and another work~\cite{n2} examined motivations behind player behaviours but did not define personalities, reducing applicability to \agentIn{'s} framework.
}

\revision{
    Additionally, we mapped high-level game entities defined by PathOS to equivalent terms in specific games. PathOS defines nine entity types in total. For example, “Enemy Hazard”, described as “A hostile character, etc., which could incite combat”, maps to “enemies” in DA and SPD, and to “mobs” in Minecraft, which uses a different term. Such mappings are straightforward to construct from game code and documentation, facilitating the extension of \agent to new games.
}



\subsubsection{Hybrid Planning to Accomplish Complex Tasks}\label{subsubsec:approach-hybrid-planner}

Many LLM-based agents generate only the next immediate action based on the current game state~\cite{Voyager2023, zhao2024thinkembodiedagentvirtual, ALFWorld20}. 
We refer to this strategy as a Bottom-Up Planner, which accomplishes tasks through individual actions.
While effective for simple tasks, this approach often fails on complex goals requiring long-horizon steps~\cite{GITM2023}. For instance, when tasked to \textit{craft a tool}, the agent may plan to \textit{collect resources} but later use them for unrelated actions, losing sight of the original goal.


We introduce the \textbf{Hybrid Planner}, which dynamically switches between Bottom-Up and Top-Down strategies to track goals and task progress better. It combines both strengths: the Bottom-Up Planner generates immediate, low-level actions, while the Top-Down Planner decomposes high-level goals into sub-plans using an LLM-based module. Each sub-plan is executed sequentially, and terminates once all sub-plans are completed or any of them is deemed infeasible.
This process helps the agent stay aligned with complex goals. The Planner begins with Bottom-Up planning. 
Once \agent completes a predefined number of tasks, it switches from the Bottom-Up to the Top-Down Planner. Subsequently, the Hybrid Planner dynamically alternates between the two modes based on \textit{plan diversity}, which is measured by tracking repeated actions and interacted objects across consecutive plans. If no new actions or objects are detected over a defined window,
the Planner switches modes to encourage more varied exploration in \agentIn{.}
To reduce hallucination where the Planner generates infeasible plans, we adopt the Prompt Chaining technique~\cite{Gadesha_Kavlakoglu_IBM_2024}, where each prompt builds on the output of the previous one to maintain contextual continuity. In our system, generated plans are verified against the game's definitions, and revision prompts are issued when misalignments are detected. This iterative process improves the feasibility and precision of the resulting plans.


\subsection{Action Summarizer}\label{subsec:approach-automatic-self-summary-and-learning-curriculum}

The LLM Action Summarizer evaluates each iteration’s execution by determining whether it successfully accomplishes the plan, and then generates a summary based on the outcome of the evaluation.
To mitigate unrealistic expectations that may misalign with the game state~\cite{marcus2020decadeaistepsrobust, huang2023survey}, we apply Prompt Chaining as introduced in~\ref{subsubsec:approach-hybrid-planner}. In this process, the Summarizer first prompts the LLM to predict the expected outcomes and game logs resulting from the action plan. Using these predictions as inputs, the Summarizer evaluates each action by comparing the inputs against the actual execution results. It then generates a reflective summary, leveraging the Chain-of-Thought (CoT) technique~\cite{ChainOfThought2023}, where a rationale accompanies every statement. 
These summaries are stored in the Memory System to inform future planning (see~\ref{subsec:approach-text-based-memory-system}).

\subsection{Action Executor}\label{subsec:approach-action-execution-module}

The Action Executor connects \agent to the game under test by translating action plans into executable forms. It supports two interface types: code snippets (\textit{Plan-to-Code}) and API input parameters (\textit{Plan-to-Parameters}).

\subsubsection{Plan-to-Code Translator}\label{subsubsec:approach-action-to-code-translator}

Some games expose control interfaces through SDKs or APIs for basic actions, requiring testers to prepare custom code scripts to assemble the APIs for complex tasks. For example, Minecraft's Mineflayer API~\cite{Mineflayer} lacks support for advanced actions, demanding extra scripting.

To address this, the Action Executor uses a Plan-to-Code Translator to convert Plans into executable code snippets that interact with game APIs. It generates reusable scripts (``Skills") based on basic API examples that \agent can invoke directly. The Action Summarizer then verifies execution against game states and logs, providing feedback to refine Skills when they fail to achieve the intended plans. This loop is essential, as LLMs often produce syntax errors, logic bugs, or infinite loops~\cite{LLMCodeGeneration2024, LLMCodeGenerationAndVerification2024}. To further address issues like infinite loops or infeasible tasks, we introduce another LLM module to allocate execution time based on plan complexity and \agentIn{'s} personality, e.g., aggressive agents allow more time for combat, while cautious ones allow less.

\subsubsection{Plan-to-Parameters Translator}\label{subsubsec:approach-action-to-parameters-translator}
Exposed APIs directly map to actions in some games. In such cases, the Action Executor translates high-level plans into API parameters, enabling seamless interaction by \agentIn{.}

\subsubsection{Custom Translators}
The Action Executor supports two translators, enabling \agent to adapt to diverse games. For games with unique architectures, developers can create custom APIs and integrate them with a well-designed Executor to bridge \agent and the game.

\subsection{Memory System}\label{subsec:approach-text-based-memory-system}

The Memory System stores actions, in-game environments, and Summaries as Memories, which are later retrieved to guide planning. These Memories help the Planner generate context-aware actions aligned with the agent's personality. However, as action iterations grow, including all Memories in prompts becomes infeasible due to token limits and increased hallucination risk~\cite{CoTFairthfulness2023, wiskojo/overwhelm-llm-eval, 1GAVillage2023}. To address this, we adopt a Retrieval-Augmented Generation (RAG) approach~\cite{11RAGOriginalPaper} to manage and retrieve only the most relevant Memories. This reduces token overhead and mitigates hallucinations~\cite{RAGSurveyGAo2024, YuHaoRAGEvaluation2024}.

\subsubsection*{\textbf{Retrieval of Preferred Plans}}

To simulate the influence of human preferences in decision-making~\cite{SimpleHeuristics1999}, each Memory is paired with a \textit{preference summary}, an LLM-generated reflection conditioned on the given personality, describing how the action and outcome preferred by such a personality. A \textit{preference score} is then computed using cosine similarity between this summary and the personality prompt, and the top five scoring Memories are retrieved as preferred plans.

\subsubsection*{\textbf{Retrieval of Related Memories}}
To mimic how human players recall past experiences~\cite{Analogy1983}, the Memory System retrieves the top five most relevant Memories based on Cosine Similarity between the current and past in-game environments. It supplies them to the Planner for the next planning phase.

\subsubsection*{\textbf{Retrieval of Skills}}
As introduced in Section~\ref{subsubsec:approach-action-to-code-translator}, reusable code snippets, called Skills, are generated to interact with games lacking SDKs or complete APIs. Each generated Skill is stored with a textual description. When generating new code, the Action Executor retrieves the top five most similar Skills based on Cosine Similarity between the current plan and the description, enabling reuse or providing references.

\section{Evaluation}\label{sec:overview-of-research-questions}

To evaluate the performance of \agentIn{,} we conducted two complementary studies: an \textit{effectiveness study} and a \textit{usefulness study} across three different game subjects. These studies address the following research questions:

\begin{itemize}[leftmargin=1.4em, itemsep=2pt, topsep=0.5pt]
    \item \textbf{RQ1}: How effective is \agent in achieving code coverage?
    \item \textbf{RQ2}: How effective is \agent in covering diverse in-game behaviours and interactions?
    \item \textbf{RQ3}: How does \agent perform compared to state-of-the-art tools in completing given tasks?
    \item \textbf{RQ4}: How diverse are \agentIn{'s} solutions in solving the given tasks compared to existing tools?
\end{itemize}

The \textit{effectiveness study} answers \textbf{RQ1} and \textbf{RQ2} by evaluating \agentIn{'s} \textit{code-} and \textit{interaction-level} coverage in two open-source games. 
The \textit{usefulness study} answers \textbf{RQ3} and \textbf{RQ4} by comparing \agent with a state-of-the-art LLM agent in \textbf{Minecraft (MC)}~\cite{Minecraft}, a widely adopted subject with rich agent baselines~\cite{61AutoMC2024}. This real-world, closed-source setting focuses on \textit{task completion} and \textit{solution diversity}, highlighting \agentIn{'s} practical performance against existing tools.

\subsection{Effectiveness Evaluation (RQ1 \& RQ2)}\label{sec:effective-evaluation}

\begin{figure}[!t]
    \centering
    \subfloat{\includegraphics[width=0.40\linewidth]{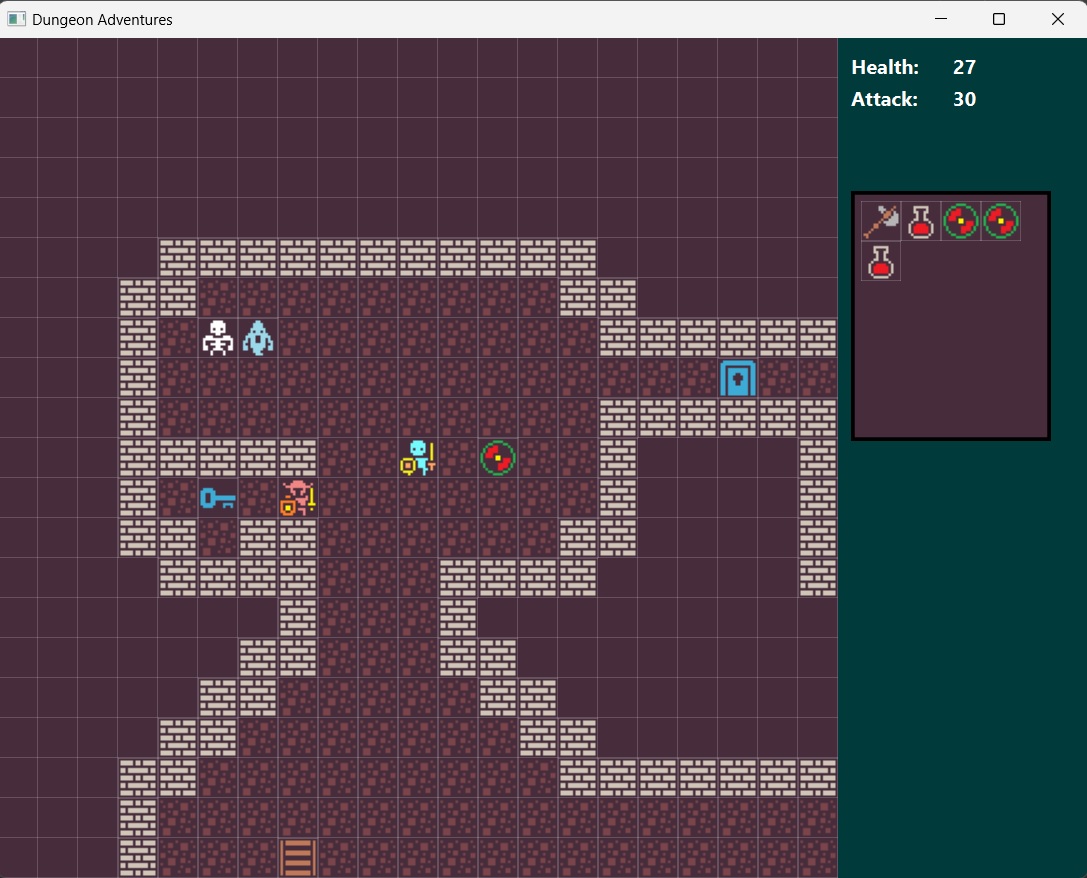}
    \label{fig:da-screenshot}}
    \hfil
    \subfloat{\includegraphics[width=0.56\linewidth]{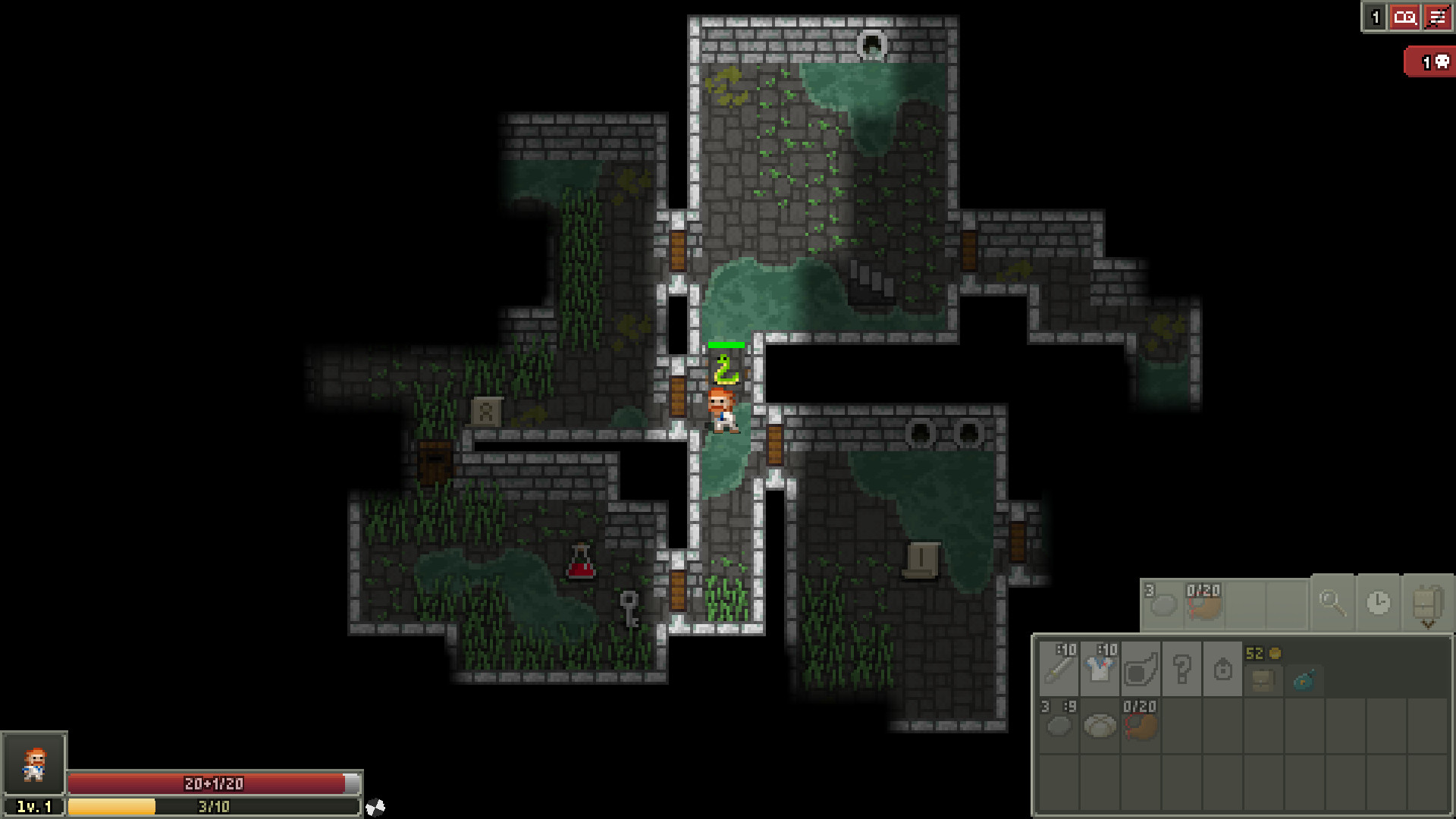}
    \label{fig:spd-screenshot}}
    \caption{Screenshots of the Dungeon Adventures (left) and Shattered Pixel Dungeon (right).}
    \label{fig:screenshot}
\end{figure}

This section evaluates \agentIn{'s} effectiveness in terms of \textit{code-level} and \textit{interaction-level} coverage using two open-source games: \textit{\textbf{Dungeon Adventures (DA)}}~\cite{DA} and \textit{\textbf{Shattered Pixel Dungeon (SPD)}}~\cite{SPD} (see screenshots in \figurename~\ref{fig:screenshot}).

\subsubsection{\textbf{Experimental Setup}}\label{subsec:effective-es}

To accommodate LLM response latency, we selected non-time-sensitive games and integrated lightweight API layers for interaction, without altering any game logic or mechanics. This preserved original gameplay for realistic, unbiased evaluation.  

DA is a small-scale, turn-based Role-Playing Game (RPG) with four levels, four item types, and four enemy types. Players can move, collect items, or engage in proximity-based combat. The gameplay is simple, with limited operations and flexibility. The testing objective for this game is to defeat the boss.

SPD is a large-scale, turn-based roguelike RPG, with randomly generated maps, items, and enemies. Since its 2014 release, SPD has 1M+ downloads and 3,900 GitHub stars. It supports complex actions, such as crafting and upgrading, with 25 levels, 250+ item types, and 65+ enemy types. The testing objective for this game is to complete the dungeon.

All LLMs in \agent used GPT-4o (version 2024-08-06)~\cite{OpenAIGPT4o} via API. Experiments were run on a 32GB RAM, 64-bit Windows 11 machine with an Intel i7-10750H CPU @ 2.60GHz (6 cores).

\subsubsection{\textbf{Baselines}} \label{subsec:effective-baselines}

We compared \agent against five baselines.

\textbf{Ablated baselines:} To assess the impact of the Memory System, personality components, and the Summarizer, we include two ablated versions of \agentIn{:} \LLMPIn{,} which contains only the LLM Planner, and \LLMPSIn{,} which adds the Summarizer but omits both the Memory System and personality modules.

\textbf{Human baselines:} Human testers serve as a manual testing baseline, a common practice in game development~\cite{SurveyVideoGameTesting}. To match the number of personalities in \agentIn{,} we recruited seven testers to play both games. 
\revision{While results from a single group of humans may not capture full variability, our sample size was constrained by budget limitations. To mitigate this threat, all testers were experienced,} each with over five years of gaming experience and averaging 6.8 hours of gameplay per week. Participants were compensated 20 CAD per hour, and no identifiable information was collected. \revision{Moreover, our results show that this group size is sufficient to highlight the performance gap between \agent and humans (see Section~\ref{threats-to-validity}).}

\textbf{Random baselines:} We do not include existing agents as baselines for RQ1-2, as they are highly tailored to specific games and cannot be easily adapted to new games. Instead, to represent automated random testing strategies, we implemented two Monkey Testing variants~\cite{RandomTestingMonkeysandGorillas, Monkey, 21Wuji}. \textbf{Dumb Monkey} randomly invokes exposed APIs with unconstrained inputs, while \textbf{Smart Monkey} ensures that all invoked actions and parameters are valid, triggering only meaningful actions.

\subsubsection{\textbf{Metrics}} \label{subsec:effective-metrics}

\subsubsection*{\textbf{Code Level Metrics}}

To assess code-level effectiveness, we measured \textit{code coverage} and \textit{branch coverage}, the de facto test coverage criteria to evaluate test cases. Since automated tools interact via APIs rather than the UI, all UI-related code, unreachable through agents, was excluded. Additionally, we excluded code modifications from our side for API development and log instrumentation, as these were inaccessible to some tools. Coverage was measured using JaCoCo~\cite{Jacoco}.

\subsubsection*{\textbf{Interaction Level Metrics}}
While code-level coverage measures how much of a game's internal logic is executed, it fails to capture how agents interact with the game world. For instance, branch coverage may confirm that an item was used. However, it cannot detect whether multiple items were used concurrently or under specific in-game conditions, which are the scenarios common in gameplay that can trigger unforeseen bugs. We propose to use \textbf{combinatorial coverage} to address this gap and evaluate diverse action-parameter combinations across conditions. Likewise, code coverage might show that a collision handler was invoked, but not the agent encountered edge cases like clipping through walls or misaligned hitboxes; such issues are more effectively assessed by \textbf{navigation coverage}, which measures the spatial exploration.
Together, these interaction-level metrics reveal behaviour-driven bugs that code-level metrics often miss.

\begin{itemize}[leftmargin=0.8em, itemsep=2pt, topsep=0.5pt]
    \item \textbf{Combinatorial Coverage:} 
    It measures the percentage of the combinations of actions and parameters explored~\cite{CombinatorialCoverageMeasurement2013, CombinatorialCoverageas2015}. To model these combinations, we define a \textit{combinatorial rule} as a tuple of an \textit{action type} (e.g., \textit{use}, \textit{throw}, \textit{eat}) and up to four parameters: subject items, targets, carrying items, and character upgrades. \textit{Subject items} are the primary objects involved in the action; \textit{targets} are entities they interact with; \textit{carrying items} and \textit{character upgrades} are boolean flags indicating inventory or acquired abilities, which may trigger special interactions. We analyzed game source code and classes to map entities to these types and to generate combinations, each representing a distinct scenario. For example, \texttt{[throw, stone, door, potion, ¬random\_upgrade]} corresponds to “the player throws a stone at a door while carrying a potion without the random upgrade”. Complete definitions are available on our project website~\cite{MIMIC_Webpage}.

    \item \textbf{Navigation Coverage:} 
    It measures how thoroughly \agent explores the game’s spatial layout. We track reachable locations across rooms and levels, then compute the proportion visited. In SPD, this includes room transitions, alternate paths, and secret areas. High navigation coverage reflects a tool's ability to uncover nonobvious paths, adapt to environmental complexity, and reveal pathfinding-related issues.
\end{itemize}

\revision{
    Although the numerical maximum of coverage is 100\%, this is rarely achievable in practice due to factors like unreachable code (e.g., dead code) and infeasible gameplay combinations. Meanwhile, precisely detecting them remains an open challenge, requiring extensive analysis and effort. Thus, the practical maximum coverage is lower. While \agent remains far from these bounds, our results show that integrating personality-driven planning into LLM-based agents improves both behavioural diversity and code coverage over existing game testing tools, highlighting ongoing gaps in automated game testing and motivating further research.
}
    
\subsubsection{\textbf{Execution Setup}} \label{subsec:effectiveness-execution-setup}

To reflect \agentIn{'s} integration of seven personalities, \textbf{one complete run consists of seven individual runs}, each corresponding to a distinct personality. All experiments were repeated thrice to account for randomness in \agentIn{'s} behaviour, resulting in 21 runs. For fairness, the same setup was applied to all other automated baselines.

For \LLMPIn{,} \LLMPSIn{,} and human testers in \textbf{SPD}, only one complete run (seven individual runs) was conducted. Since these baselines do not incorporate personality variations, a single complete run with seven repetitions sufficiently accounts for their behavioural randomness. Our evaluation confirms that this reduced run count did not impact the reliability of performance comparisons: human testers generally outperformed automated tools, while \LLMP frequently stalled within the first hour, and \LLMPS consistently achieved lower final coverage than the complete version of \agentIn{.}

\begin{itemize}[leftmargin=0.8em, itemsep=2pt, topsep=0.5pt]
    \item \textbf{Setup for DA}: Automated tools were given a time limit of one and a half hours per run. Based on observed efficiency, human testers were allocated one hour per run, as their coverage typically converged more quickly and extended playtime offered diminishing returns.
    The game was modified to automatically restart upon player death or victory, with no other changes to its original design.
    \item \textbf{Setup for SPD}: All tools and testers were given a four-hour time limit per run. The game auto-restarted after each death or victory, generating a new map using a predefined list of random seeds to ensure consistency across SPD's roguelike environment for different tools.
\end{itemize}

\subsubsection{\textbf{Code Level Effectiveness (RQ1)}} \label{subsec:effective-code-cov}

\begin{figure}[!t]
    \centering
    \subfloat{\includegraphics[width=0.48\linewidth]{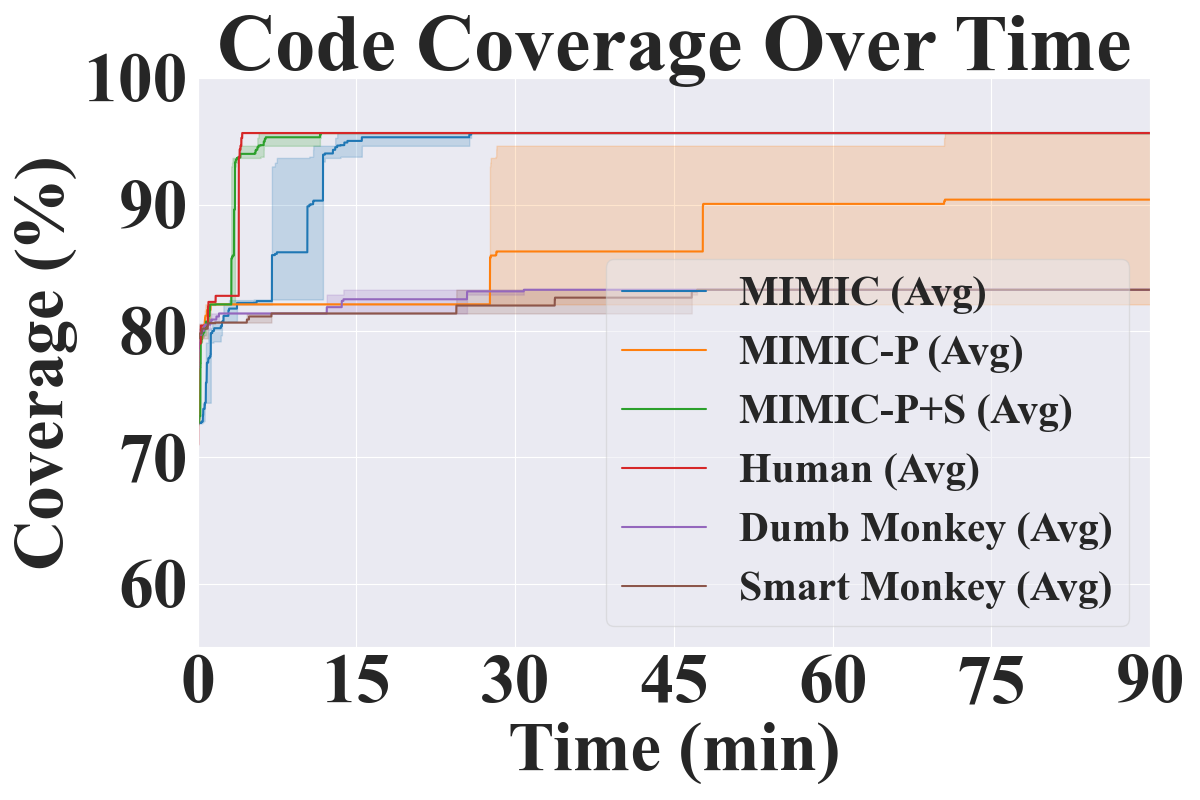}
    \label{fig:da-instruction-cov}}
    \hfil
    \subfloat{\includegraphics[width=0.48\linewidth]{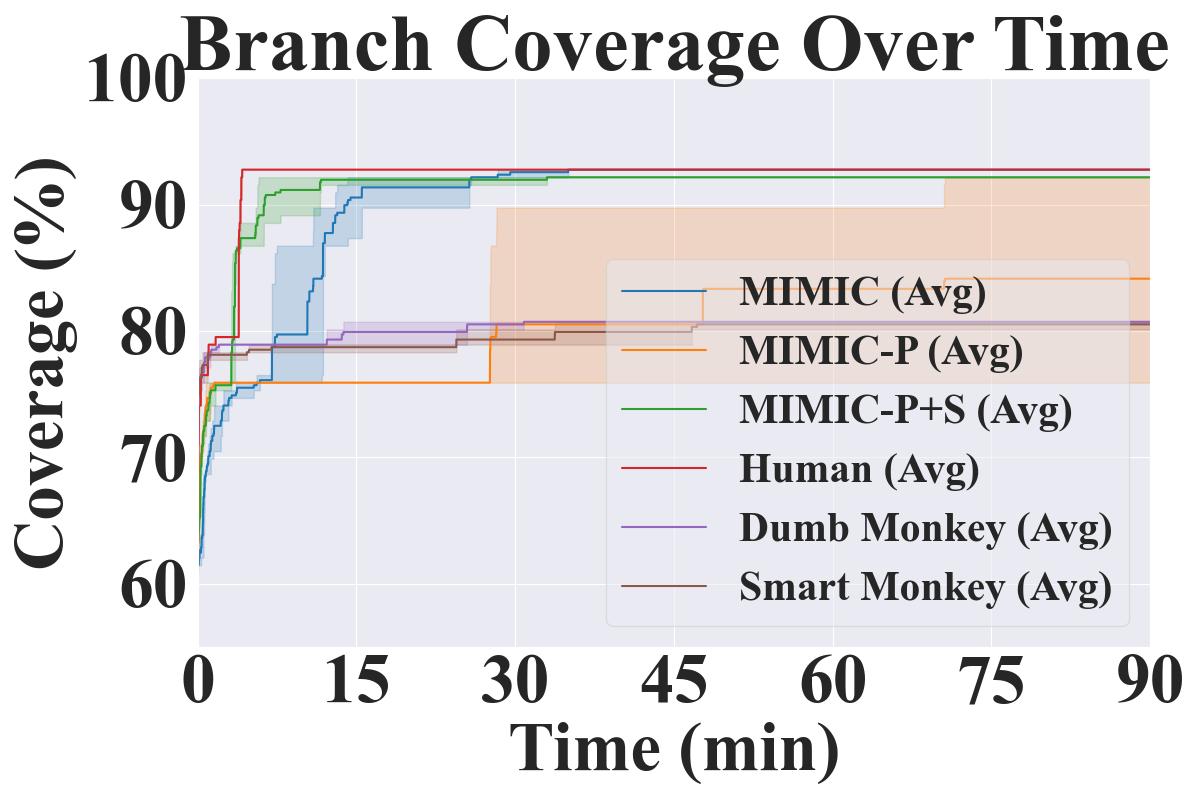}
    \label{fig:da-branch-cov}}
    \caption{Code coverage (left) and branch coverage (right) for \textbf{Dungeon Adventures (DA)}. The shaded areas represent the range across three runs, while the solid lines indicate the average coverage over time. The human completed only one complete run (no shaded area).}
    \label{fig:DA_code_results}
\end{figure}

\begin{figure}[!t]
    \centering
    \subfloat{\includegraphics[width=0.48\linewidth]{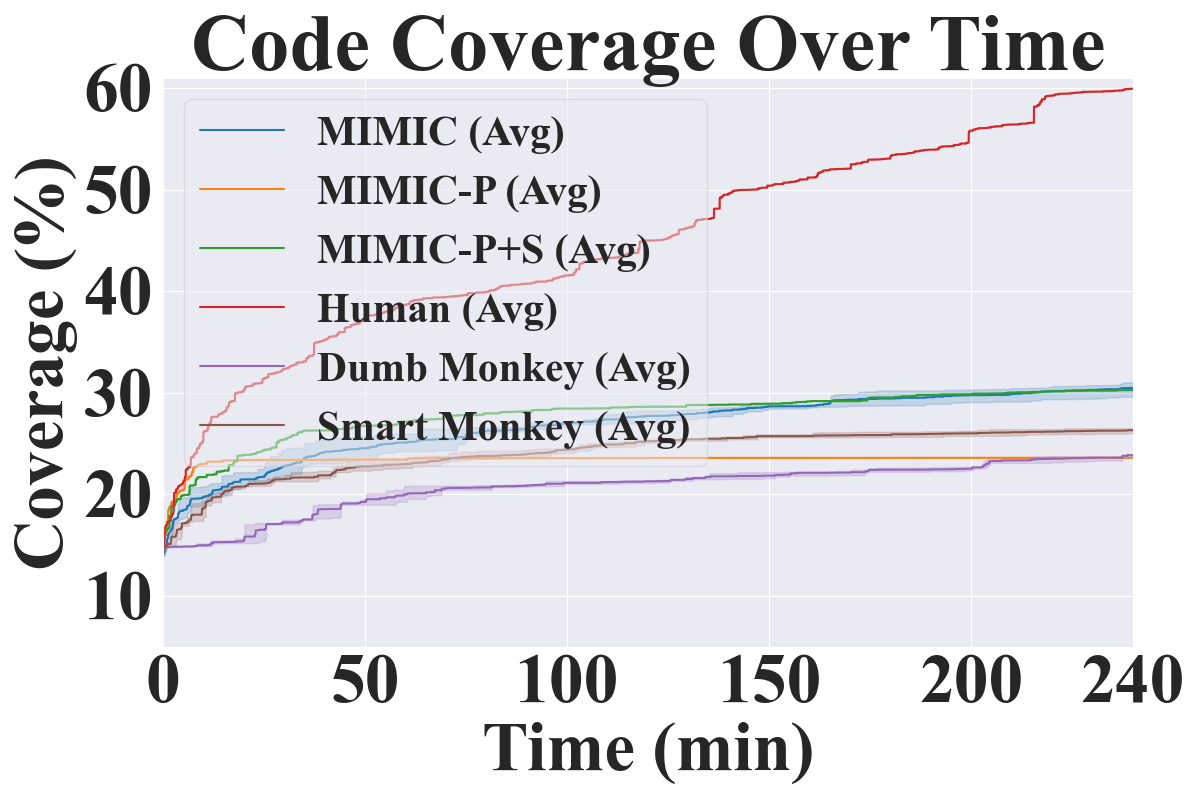}}
    \hfil
    \subfloat{\includegraphics[width=0.48\linewidth]{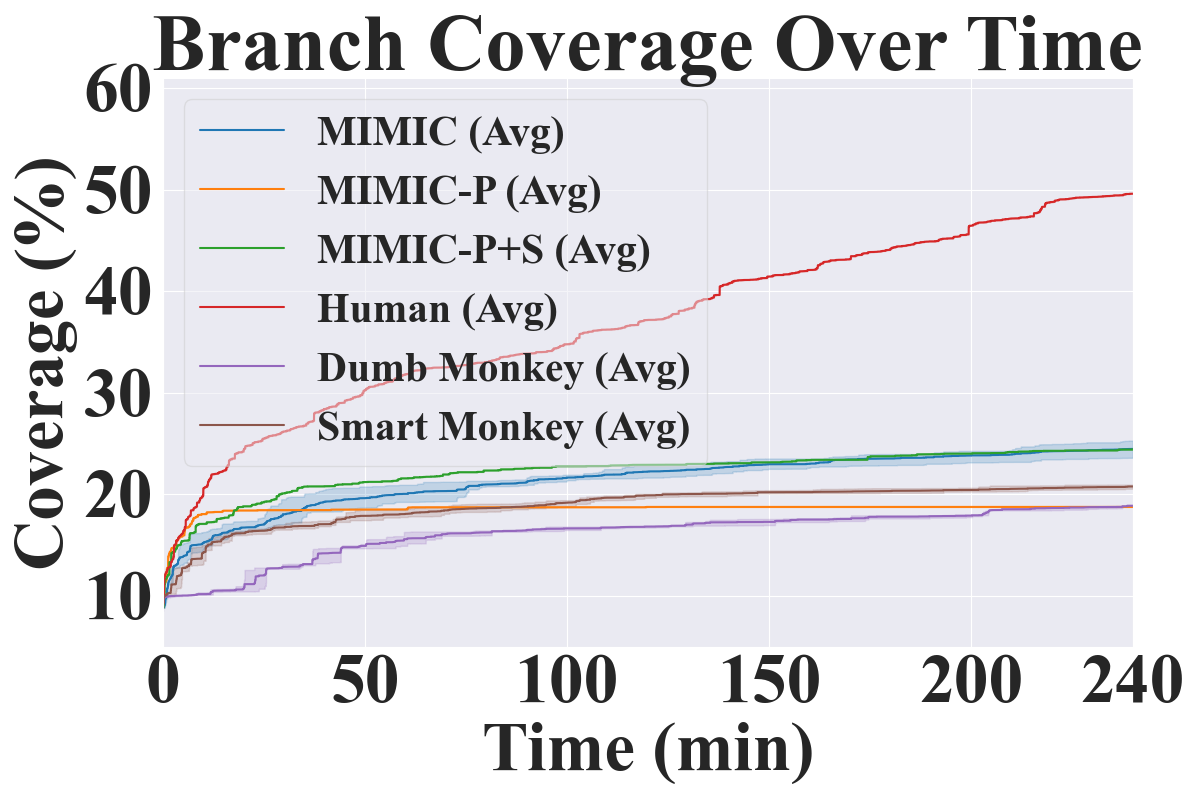}}
    \caption{Code Coverage (left) and Branch Coverage (right) for \textbf{Shattered Pixel Dungeon (SPD)} Over Time. \LLMPIn{,} \LLMPSIn{,} and Human completed only one complete run (no shaded area).}
    \label{fig:SPD_code_results}
\end{figure}

\figurename~\ref{fig:DA_code_results} and \figurename~\ref{fig:SPD_code_results} show code and branch coverage over time in DA and SPD, respectively. In DA, \agent achieved the highest code (95.67\%) and branch (92.77\%) coverage within 35 minutes. In SPD, despite the game's larger scale and complexity, \agent still outperformed all automated tools, reaching 30.50\% code and 24.49\% branch coverage.

Compared to humans and Monkeys, \agent demonstrated superior performance. While its early progress in DA was slightly slower than that of human testers, it eventually matched their final coverage levels with a higher branch coverage. Against the Monkeys, \agent achieved 1.16$\times$ higher coverage in both code and branches. In SPD, \agent achieved 3\% higher coverage on average than Smart Monkey, corresponding to 11,776 additional lines and 1,115 more branches, and 6\% higher coverage than Dumb Monkey.

\begin{figure}[!t]
    \centering
    \subfloat{\includegraphics[width=0.48\linewidth]{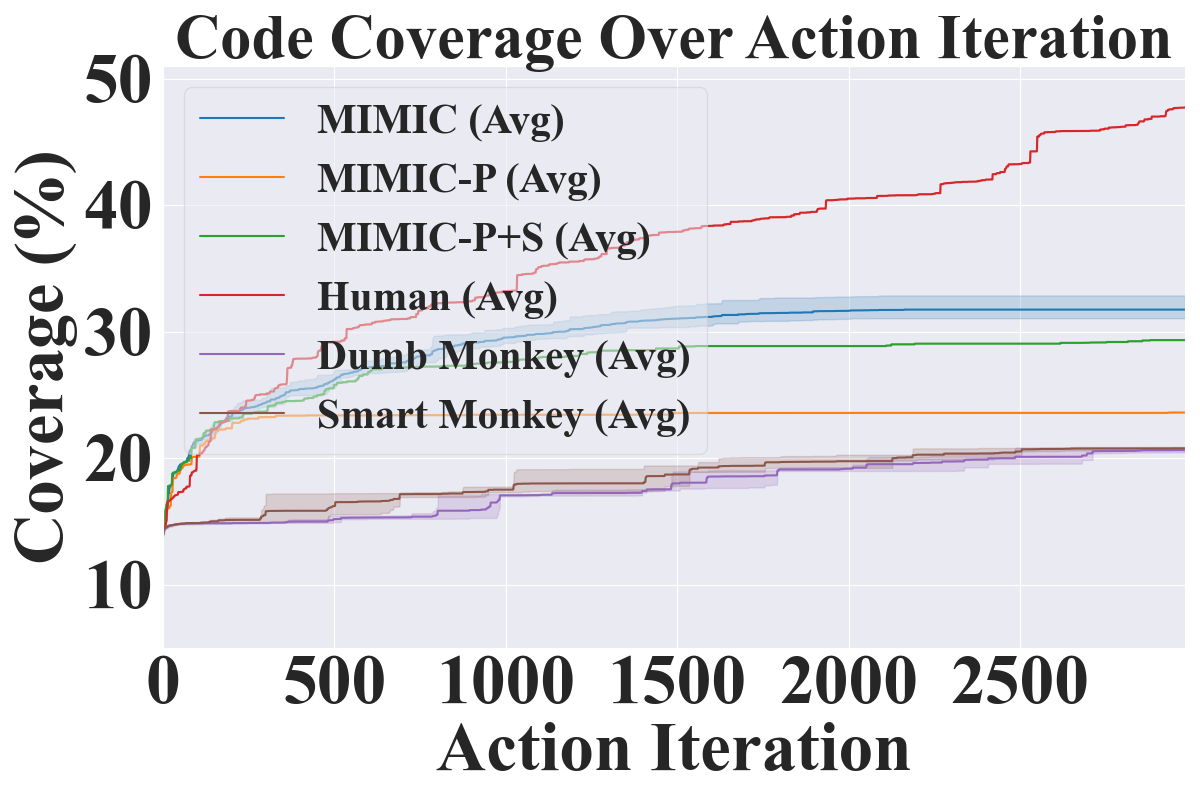}}
    \hfil
    \subfloat{\includegraphics[width=0.48\linewidth]{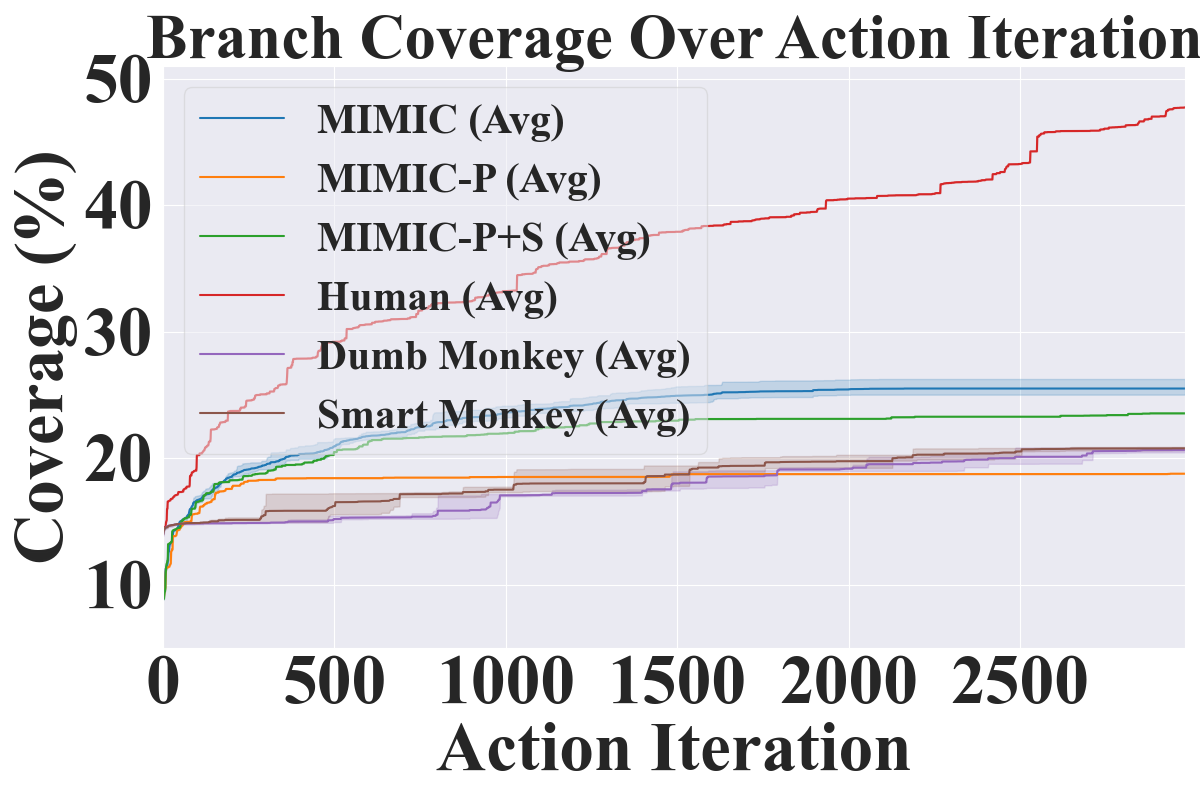}}
    \caption{Code Coverage (left) and Branch Coverage (right) for \textbf{Shattered Pixel Dungeon (SPD)} Over Action Iteration.}
    \label{fig:SPD_code_results_over_action}
\end{figure}

Compared to ablated versions, \agent consistently achieved the best results. In DA, \agent achieved 1.06$\times$ higher code and branch coverages than \LLMPIn{,} and ultimately outperformed \LLMPS in branch coverage. In SPD, \agent achieved 6\% higher average coverage than \LLMP across both metrics. Although the improvement over \LLMPS was minor, with 0.21\% in code and 0.04\% in branch coverage, this is mainly due to action throughput differences: \LLMPS executed around 2,500 actions per run with two LLM components, while \agent executed only 1,500 with three to four LLM components, including memory retrieval. To better demonstrate the impact of the throughput differences, \figurename~\ref{fig:SPD_code_results_over_action} plots coverage per action iteration, where \agent consistently achieved higher coverage efficiency, outperforming \LLMPS by around 2\% in coverage per action.
These results confirm that all components in \agent holistically contribute to the performance. 
\LLMPSIn{'s} improvement over \LLMP underscores the Summarizer's role in maintaining planning context.
With \agent outperforming \LLMPS with even fewer actions, Memory System and personality-driven planning are proven to be effective in driving more diverse and effective gameplay exploration. 

Currently, \agent uses ChatGPT-4o via the OpenAI API, which introduces communication overhead and is not dedicated to game interaction or personality mimicking. Despite this, \agent still shows effectiveness in achieving high code-level coverage. Future work will explore locally fine-tuned LLMs tailored to game-specific tasks, which could further enhance efficiency and better align the system with the demands of automated game testing.


\subsubsection{\textbf{Interaction Level Effectiveness (RQ2)}} \label{effective-interaction-cov}

\begin{figure}[!t]
    \centering
    \subfloat[Combinatorial Coverage in DA]{\includegraphics[width=0.48\linewidth]{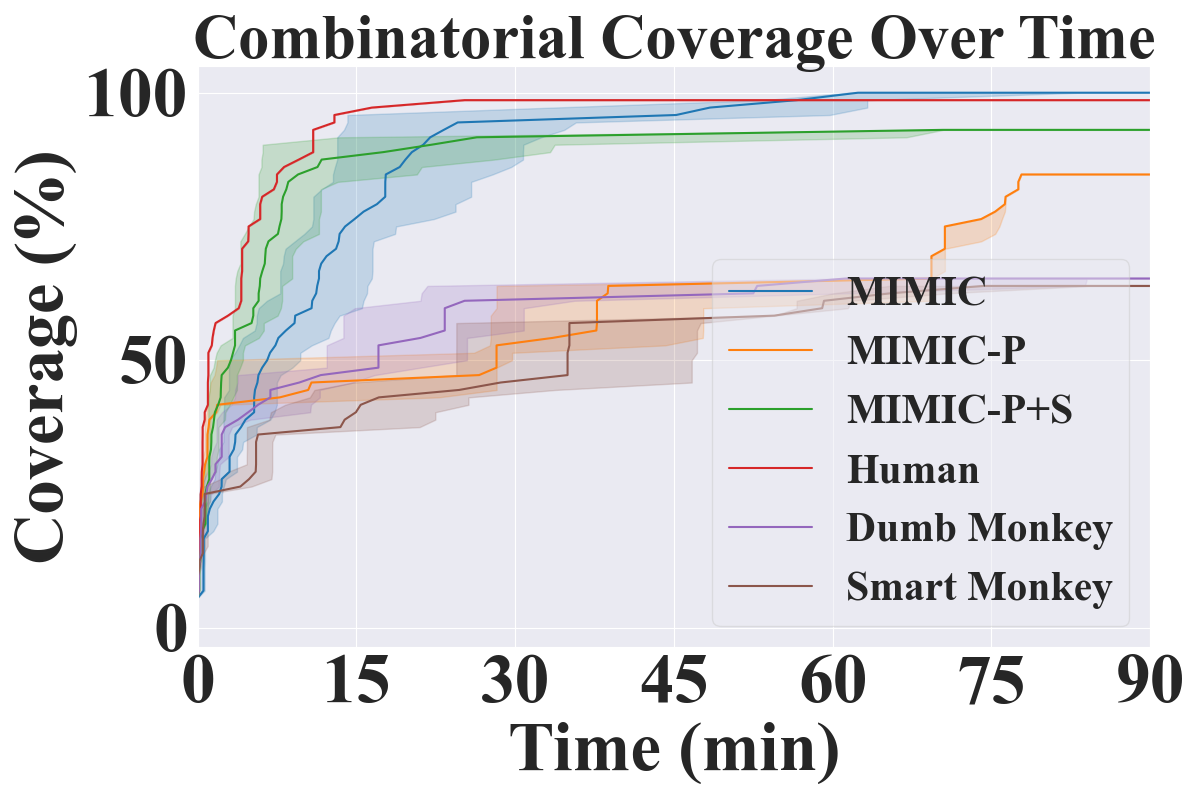}
    \label{fig:DA_comb_coverage}}
    \hfil
    \subfloat[Combinatorial Coverage in SPD]{\includegraphics[width=0.48\linewidth]{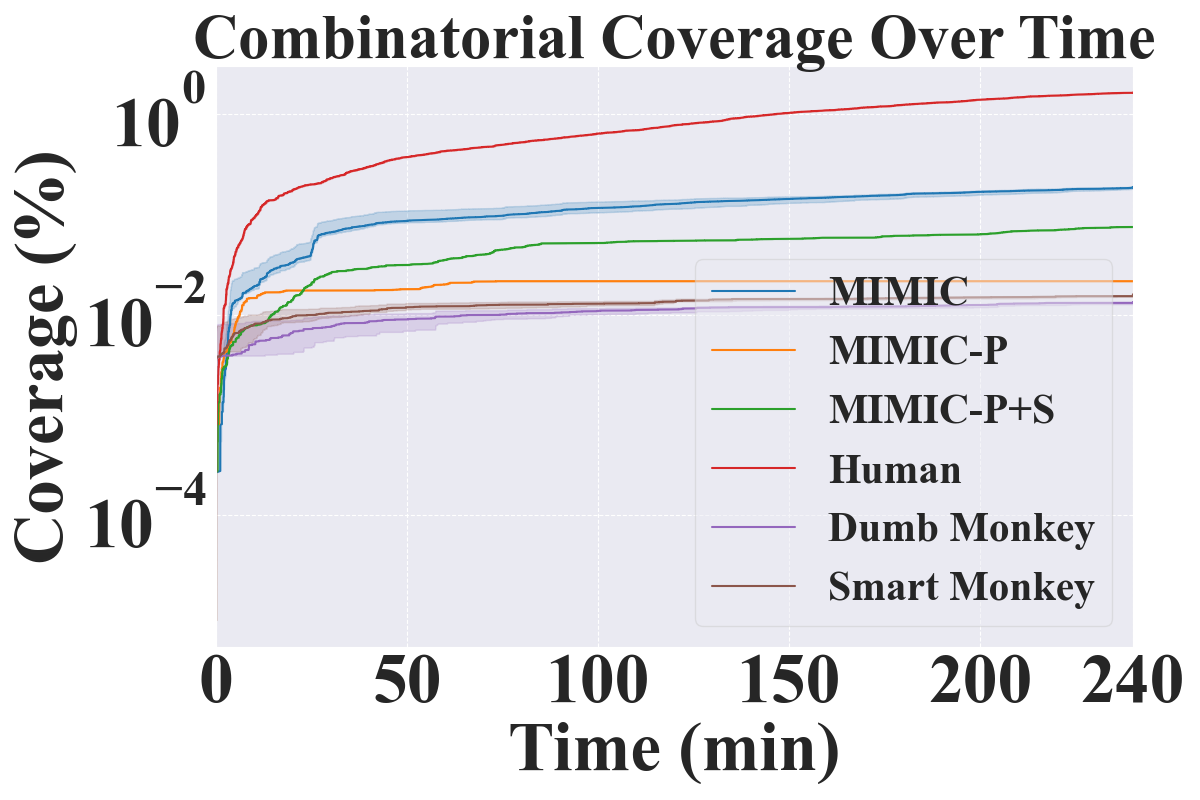}}
    \label{fig:SPD_comb_coverage}
    \caption{ Combinatorial coverage for \textbf{Dungeon Adventures (DA)} (left) and \textbf{log-scaled} coverage for \textbf{Shattered Pixel Dungeon (SPD)} (right). The shaded areas represent the range across three runs, while the solid lines indicate the mean time to cover individual combinations. }
\end{figure}

\agent demonstrates strong effectiveness over automated tools in exploring diverse and meaningful gameplay interactions, as reflected in both combinatorial and navigation coverage metrics. In DA, \agent achieves 100\% combinatorial coverage across 72 defined combinations (\figurename~\ref{fig:DA_comb_coverage}) and records the highest average navigation coverage among automated tools (Table~\ref{tab:level-exploration}). In SPD, despite its significantly larger, procedurally generated environment, \agent attains the highest combinatorial coverage among automated tools, covering 0.188\% (21,319 out of 11.3 million combinations, \figurename~6b), and reaches the deepest average navigation levels and highest averaged navigation coverage among automated tools (Table~\ref{tab:level-exploration}). For consistency, navigation coverage in SPD is reported over the first four levels due to its random seeding of large-scale maps, which also aligns with the typical exploration range across tools. These results demonstrate \agentIn{'s} ability to navigate and interact within complex and large-scale game environments.

Compared to human testers, \agent achieved lower navigation coverage but eventually surpassed them in combinatorial coverage in DA. In both games, human testers maintained an edge in level progression and navigation coverage. However, relative to Dumb Monkey and Smart Monkey, \agent showed consistent advantages: in DA, it achieved 1.5$\times$ higher combinatorial coverage; in SPD, it covered 2.51$\times$ more combinations than Smart Monkey and reached significantly greater navigation depth and coverage. Although \agent did not outperform human testers in SPD, its continuous improvement without saturation over four hours underscores its capacity for long-term exploration and interaction learning.

\agent also consistently outperformed its ablated variants. In DA, it achieved 1.18$\times$ higher combinatorial coverage than \LLMP and 1.07$\times$ more than \LLMPSIn{.} In SPD, \agent reached 0.188\% coverage, substantially outperforming \LLMP (0.022\%) and \LLMPS (0.075\%), which is especially significant given the scale of the combination space. In navigation, \agent explored both more deeply and with higher averaged coverage than its ablations (1.28$\times$-1.35$\times$ higher). These gains highlight the role of memory retrieval and personality-driven planning in promoting more varied, goal-aware exploration.

To reduce bias from predefined rules, we also measured coverage over all interactable object types (e.g., terrain, characters, items), observing consistent trends. This confirms the value of interaction-level metrics and \agentIn{'s} effectiveness in testing complex gameplay scenarios. 

\begin{table}[!t]
    \caption{\textbf{Average levels explored \& average navigation coverage across the map by different tools.} 
    }
    \label{tab:level-exploration}
    \centering

    \resizebox{\columnwidth}{!}{
    \begin{tabular}{lccc}
    \toprule
        \textbf{Game} & \textbf{Testing Tool} & \textbf{Avg. Lvl. Explored ± Std.} & \textbf{Avg. Nav. Cov. ± Std.} \\ 
        \midrule
        \multirow{6}{*}{\textbf{DA}} 
        & \agent & 3.00 ± 1.13 & 52.16 ± 5.56 \\
        & \LLMP & 2.00 ± 1.32 & 23.62 ± 9.13 \\
        & \LLMPS & 3.07 ± 1.39 & 45.67 ± 1.93 \\
        & Human & 2.43 ± 0.87 & 99.45 ± N/A \\ 
        & Dumb Monkey & 1.93 ± 0.26 & 32.08 ± 2.88 \\
        & Smart Monkey & 1.71 ± 0.47 & 30.22 ± 3.72 \\  \midrule

        \multirow{6}{*}{\textbf{SPD}}
        & \agent & 2.28 ± 0.92 & 14.41 ± 7.51 \\
        & \LLMP & 1.90 ± 0.66 & 11.27 ± 5.77 \\
        & \LLMPS & 1.94 ± 0.82 & 10.67 ± 4.92 \\
        & Human & 3.89 ± 2.90 & 47.07 ± 20.29 \\ 
        & Dumb Monkey & 1.07 ± 0.27 & 5.75 ± 3.15 \\
        & Smart Monkey & 1.05 ± 0.24 & 5.77 ± 3.49 \\ 
        
    \bottomrule
    \end{tabular}}
    
\end{table}

\revision{
    Since only one complete run was conducted for the human group in both games, statistical tests are not appropriate, as they require multiple samples to estimate population parameters.
    Instead, we report confidence intervals for these two groups in Table~\ref{tab:CIs}. In DA, the intervals are close across all three metrics, showing comparable performance with \agent achieving marginal improvements. In SPD, a more complex environment, the human group achieves substantially higher coverage with nonoverlapping intervals, highlighting the significant performance gap between humans and \agentIn{.}
}

\subsection{Usefulness Evaluation (RQ3 \& RQ4)}\label{sec:comparison-evaluation}

This section evaluates \agentIn{'s} usefulness by comparing it to a state-of-the-art LLM-based agent in \textbf{Minecraft (MC)}, focusing on two dimensions: \textit{task completion} (RQ3) and \textit{solution diversity} (RQ4). These are measured by success rates and variation in task-solving interactions, respectively, to assess \agentIn{'s} performance relative to existing tools.

\subsubsection{\textbf{Experimental Setup}}\label{subsec:comparison-es}

MC is selected as the subject for this study due to its rich action space and open-ended gameplay, making it widely used for gaming agent evaluation~\cite{61AutoMC2024}. Its popularity has led to the development of many LLM-based agents, enabling meaningful comparisons. 

\begin{table}[!t]
    \caption{\revision{\textbf{95\% Confidence Intervals of Coverage for MIMIC and Human.} 
    Note that the Human baseline includes only one complete run, so the interval equals the observed average.}}
    \label{tab:CIs}
    \centering
    \resizebox{\columnwidth}{!}{
    \revision{\begin{tabular}{lcccc}
      \toprule
        \textbf{Game} & \textbf{Tool} & \textbf{Code} & \textbf{Branch} & \textbf{Combinatorial} \\
            \midrule
            \multirow{2}{*}{\textbf{DA}} 
            & MIMIC & [95.7\%, 95.7\%] & [93.06\%, 93.06\%] & [92.88\%, 100.0\%] \\
            & Human* & [94.94\%, 94.94\%] & [89.02\%, 89.02\%] & [98.61\%, 98.61\%] \\  \midrule
    
            \multirow{2}{*}{\textbf{SPD}} 
            & MIMIC & [28.62\%, 32.37\%] & [22.38\%, 26.59\%] & [0.105\%, 0.271\%] \\
            & Human* & [59.97\%, 59.97\%] & [49.63\%, 49.63\%] & [1.646\%, 1.646\%] \\
        \bottomrule
    \end{tabular}}}
\end{table}

We compared with \textbf{ODYSSEY}~\cite{56Odyssey2024} \revision{in its original design} as the baseline since it's a state-of-the-art gaming agent for MC with strong problem-solving performance (e.g., achieving over 90\% success on the challenging \textit{Obtain 1 Diamond} task).
It consistently outperforms other existing agents, including Voyager~\cite{Voyager2023}, GITM~\cite{GITM2023}, VPT~\cite{48VPT2022}, and DEPS~\cite{81DEPS}. ODYSSEY comprises a Planner, an Action Executor, and a Critic for summarizing plans. 
Unlike \agentIn{'s} code-generation approach, its Executor selects from a library of 183 pre-coded functions, ranks the top ten via Cosine Similarity, and uses an LLM to invoke the most relevant one. 
ODYSSEY uses MineMA-8B and MineMA-70B, both fine-tuned LLaMA-3 models~\cite{MInEMA}. 
Following its original evaluation, we use MineMA-8B in all ODYSSEY experiments, where it was used in most tasks.

\revision{To preserve the integrity of the baseline and ensure a fair comparison, we evaluated ODYSSEY strictly with its original skill functions. While combining our Plan-to-Code Translator with ODYSSEY may improve success rates by generating richer skills, it would not increase the solution diversity. 
This limitation stems from ODYSSEY's LLM planner, which lacks personalities to generate varied solutions for the same goals. Replacing the code-generation component does not address this limitation.
Moreover, removing the hardcoded skills would effectively make ODYSSEY similar to our \LLMPS baseline (identical to \agentIn{,} but without personalities; see Section~\ref{subsec:effective-baselines}), which already underperforms \agent in solution diversity (Section~\ref{subsec:effective-code-cov} and \ref{effective-interaction-cov}).}

Since MC is closed-source, \agent interacts with the game using its Plan-to-Code Translator (Section~\ref{subsubsec:approach-action-to-code-translator}) to generate Skills via Mineflayer~\cite{Mineflayer}. All LLM components in \agent use GPT-4o (version 2024-08-06)~\cite{OpenAIGPT4o}, accessed via API calls. All experiments were conducted on a machine with 128 GB unified memory, running macOS Sequoia 15.2 and powered by an Apple M3 Max processor with 16 cores.

\subsubsection{\textbf{Evaluation Setup}}

Unlike previously tested games, MC is not level-based and lacks predefined milestones. For fair comparison, we adopted ODYSSEY's task suite, selecting eight diverse tasks within budget limits. Each task was executed on three randomly selected maps, with three complete runs per tool (each comprising seven individual runs). For \agentIn{,} each run included all seven personalities.

Tasks were categorized by the MineDojo benchmark~\cite{60MineDojo2022} into four groups: \textit{combat}, \textit{tech tree}, \textit{harvest}, and \textit{survive}. To capture varying complexity, we sampled ODYSSEY tasks, referring to the minimum, median, and maximum number of action iterations required for ODYSSEY to achieve, and added two long-horizon tasks: \textit{Obtain 1 Diamond} and \textit{Survive 1 Day}. The final suite is summarized in Table~\ref{tab:task_suite}, grouped into two categories for clarity:

\begin{itemize}[leftmargin=0.8em, itemsep=2pt, topsep=0.5pt]
    \item \textbf{Goal-Driven Tasks}: These tasks are under the \textit{harvest} category with clear goal to collect a specific item. Each task had a one-hour time limit, except \textit{Obtain 1 Diamond}, which was allocated two hours due to its complexity.
    
    \item \textbf{Time-Limited Tasks}:
    These tasks emphasize performance within a fixed time window (i.e., one in-game day) rather than completing a specific collection goal, including all \textit{combat} tasks and \textit{Survive 1 Day}. For combat tasks, agents were teleported to a battle arena after one day, and success was defined by defeating a specified creature.

\end{itemize}

\subsubsection{\textbf{Task Completion (RQ3)}} \label{subsec:results-task-completion}

To evaluate \agentIn{'s} effectiveness in completing in-game tasks, we measured success rate within a fixed time budget and average completion time. As described in Section~\ref{subsec:comparison-es}, each complete run comprises seven individual runs. For fairness in the RQ4 diversity comparison, we conducted additional ODYSSEY runs when its total runtime was shorter than \agentIn{'s}.

\subsubsection*{\textbf{Goal-Driven Tasks}}

\agent achieved a 100\% success rate across all tasks, including the more complex ones, as shown in Table~\ref{tab:goal-driven-tasks}. In contrast, ODYSSEY succeeded in only 10 out of 21 runs for \textit{Shear 1 Sheep} and just 2 out of 21 runs for \textit{Cook 1 Meat}, consistent with its original performance reports. The superior performance of \agent is mainly attributed to \agentIn{'s} Hybrid Planner. Unlike ODYSSEY's Top-Down Planner, which strictly follows hierarchical decomposition, \agent can dynamically fall back to immediate, low-level planning when high-level subgoals repeatedly fail. This flexibility enables \agent to adapt its strategy in response to execution failures, improving robustness and task completion in complex or unpredictable scenarios.


In terms of efficiency, \agent outperformed ODYSSEY in most cases, except for the most challenging task, \textit{Obtain 1 Diamond}, where ODYSSEY completed all runs significantly faster than \agentIn{.}  However, its inconsistent performance on simpler tasks suggests it may be overfitting to this particular benchmark. To understand \agentIn{'s} longer completion time in \textit{Obtain 1 Diamond}, we analyzed the performance across its seven personality-driven agents. The \textit{aggressive} personality emerged as an outlier, averaging 102.44 minutes, while others completed the task in 25–35 minutes. Further inspection revealed that aggressive agents often prioritized combat over task progression, spending more time preparing for battles unrelated to the task, consistent with their defined personality traits.
This demonstrates \agentIn{'s} ability to emulate diverse player types and realistic human decision-making.

\begin{table}[!t]
    \caption{\textbf{Task suite used for the comparison.} 
    }
    \label{tab:task_suite}
    \centering
    \resizebox{0.48\textwidth}{!}{
    \begin{tabular}{lccc}
        \toprule
        \textbf{Task ID} & \textbf{Task} & \textbf{Complexity} & \textbf{Category} \\
        \midrule
        GD\#1 & Make 1 Sugar & 1 & Tech Tree + Harvest \\
        GD\#2 & Shear 1 Sheep & 2 & Tech Tree + Harvest \\
        GD\#3 & Cook 1 Meat & 3 & Tech Tree + Harvest \\
        GD\#4 & Obtain 1 Diamond & 4 & Harvest \\
        TL\#1 & Combat 1 Cave Spider & 1 & Combat \\
        TL\#2 & Combat 1 Skeleton & 2 & Combat \\
        TL\#3 & Combat 1 Spider & 3 & Combat \\
        TL\#4 & Survive 1 Day & 4 & Survival \\
        \bottomrule
    \end{tabular}
    }

\end{table}

\begin{table}[!t]

    \centering
    \caption{\textbf{Performance comparison of different agents on the goal-driven tasks.} ``Time (min)" refers to the average minutes spent in completing the tasks. All evaluations are only calculated for successful tasks. ``±" indicates one standard deviation of the average evaluation over successful tasks. }
    \label{tab:goal-driven-tasks}

    \resizebox{0.48\textwidth}{!}{
    \begin{tabular}{llccc}
        \toprule
        \textbf{Task} & \textbf{Map} & \textbf{Group} & \textbf{Success Rate (\%)} & \textbf{Time (min)} \\ 
        \midrule
        \multirow{6}{*}{Make 1 Sugar} 
        & \multirow{2}{*}{Map 1} & \agent & 7 / 7 (100.0) & \textbf{7.55 ± 2.67}  \\ 
        &                        & ODYSSEY & 7 / 7 (100.0) & 10.6 ± 8.43  \\  \cmidrule(lr){2-5}
        & \multirow{2}{*}{Map 2} & \agent & 7 / 7 (100.0) & \textbf{7.9 ± 3.86}  \\ 
        &                        & ODYSSEY & 7 / 7 (100.0) & 8.61 ± 2.11  \\  \cmidrule(lr){2-5}
        & \multirow{2}{*}{Map 3} & \agent & \textbf{7 / 7 (100.0)} & 15.2 ± 10.79  \\ 
        &                        & ODYSSEY & 10 / 11 (90.90) & \textbf{7.7 ± 5.77} \\  \midrule

        \multirow{6}{*}{Shear 1 Sheep} 
        & \multirow{2}{*}{Map 1} & \agent & \textbf{7 / 7 (100.0)} & \textbf{17.68 ± 6.28}  \\ 
        &                        & ODYSSEY & 3 / 7 (42.86) & 28.64 ± 18.03  \\  \cmidrule(lr){2-5}
        & \multirow{2}{*}{Map 2} & \agent & \textbf{7 / 7 (100.0)} & 21.69 ± 3.25 \\ 
        &                        & ODYSSEY & 3 / 7 (42.86) & \textbf{16.75 ± 4.85} \\  \cmidrule(lr){2-5}
        & \multirow{2}{*}{Map 3} & \agent & \textbf{7 / 7 (100.0)} & \textbf{22.79 ± 6.54} \\ 
        &                        & ODYSSEY & 4 / 7 (57.14) & 38.11 ± 11.1 \\  \midrule

        \multirow{6}{*}{Cook 1 Meat} 
        & \multirow{2}{*}{Map 1} & \agent & \textbf{7 / 7 (100.0)} & \textbf{15.95 ± 7.24} \\ 
        &                        & ODYSSEY & 0 / 7 (0.0) & N/A \\  \cmidrule(lr){2-5}
        & \multirow{2}{*}{Map 2} & \agent & \textbf{7 / 7 (100.0)} & \textbf{11.86 ± 9.68} \\ 
        &                        & ODYSSEY & 1 / 7 (14.29) & 16.86 ± 0.0 \\  \cmidrule(lr){2-5}
        & \multirow{2}{*}{Map 3} & \agent & \textbf{7 / 7 (100.0)} & \textbf{16.25 ± 6.08} \\ 
        &                        & ODYSSEY & 1 / 7 (14.29) & 26.01 ± 0.0 \\  \midrule

        \multirow{6}{*}{Obtain 1 Diamond} 
        & \multirow{2}{*}{Map 1} & \agent & 7 / 7 (100.0) & 53.49 ± 56.69 \\ 
        &                        & ODYSSEY & 64 / 64 (100.0) & \textbf{6.01 ± 2.14} \\  \cmidrule(lr){2-5}
        & \multirow{2}{*}{Map 2} & \agent & 7 / 7 (100.0) & 34.04 ± 23.38 \\ 
        &                        & ODYSSEY & 29 / 29 (100.0) & \textbf{8.26 ± 2.97} \\  \cmidrule(lr){2-5}
        & \multirow{2}{*}{Map 3} & \agent & 7 / 7 (100.0) & 37.77 ± 11.09 \\ 
        &                        & ODYSSEY & 23 / 23 (100.0) & \textbf{11.69 ± 5.05} \\  
        \bottomrule
        
    \end{tabular}
    }

\end{table}

These results show that \agent not only maintains high task completion across varying complexity levels but also consistently reflects its consistency with defined personality traits, resulting in more realistic actions over speed optimization.

\subsubsection*{\textbf{Time-Limited Task}}
With a fixed time budget for time-limited tasks, we compare only success rates. Both \agent and ODYSSEY achieved 100\% across all tasks.

\subsubsection{\textbf{Task Solution Diversity (RQ4)}} \label{subsec:task-solution-diversity}

To analyze the diversity of task solutions, we collected all actions performed during task completion. We computed Shannon Entropy~\cite{Shannon_1948}, treating each action as an individual data point to quantify variability in agent behaviour. Recognizing that the sequence of actions can also influence the game, we extended this analysis by applying n-gram-based Shannon Entropy~\cite{n_grams1995, n_grams_Shannon}. It treats sub-sequences of $n$ consecutive actions as individual data points, providing a more nuanced evaluation of solution diversity.
Higher Shannon Entropy indicates greater diversity.

\begin{figure}[!t]
    \centering
    \includegraphics[width=\linewidth]{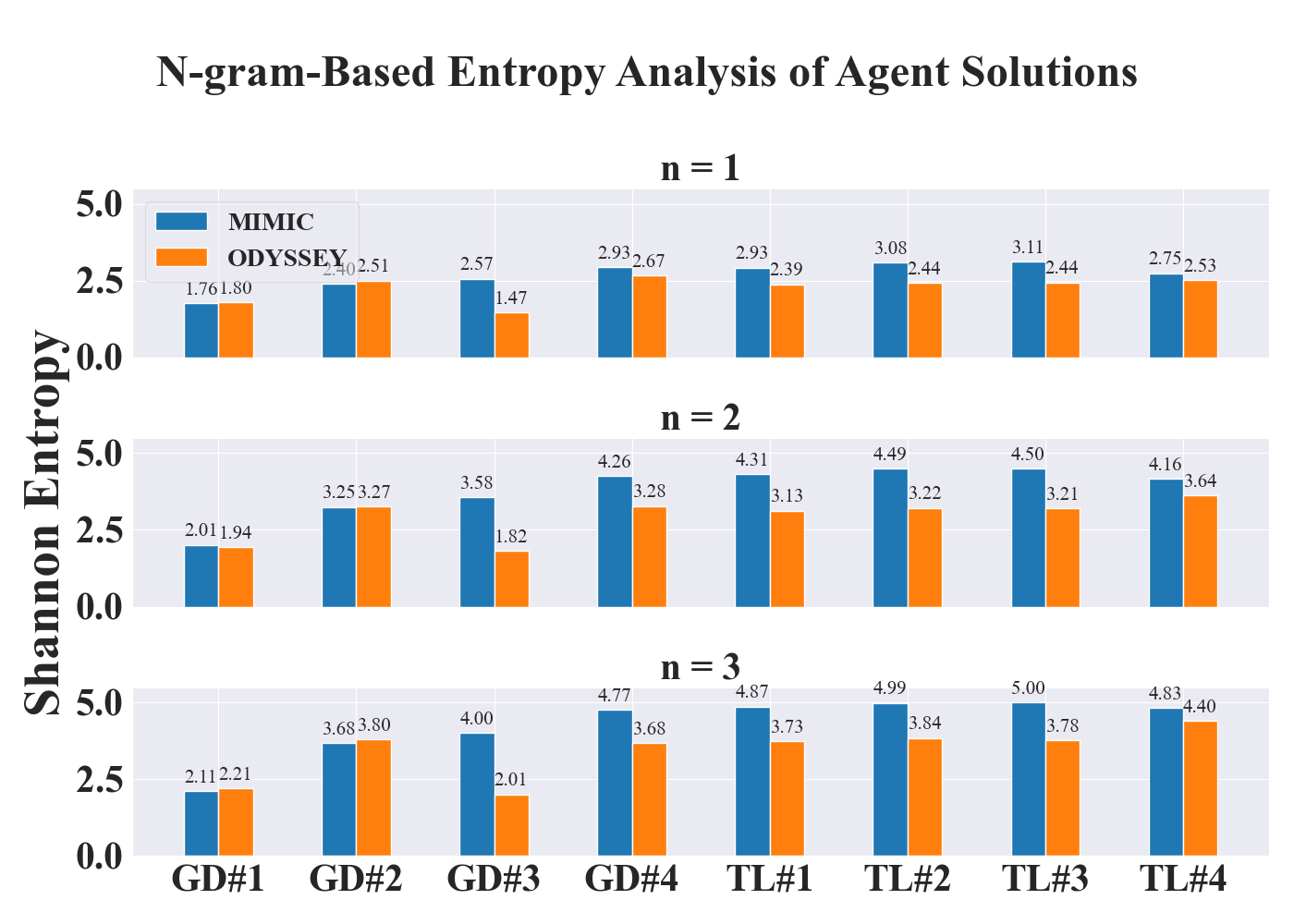}
    \caption{Average Shannon Entropy of solutions generated by different agents in solving Minecraft tasks. Results are shown separately for N-grams with varying values of $n$. Each label stands for a task with an ID given in Table~\ref{tab:task_suite}.}
    
    \label{fig:MC-diversity}
\end{figure}

As shown in \figurename~\ref{fig:MC-diversity}, \agent and ODYSSEY perform similarly on simple goal-driven tasks (\textit{Make 1 Sugar} and \textit{Shear 1 Sheep}), which is expected since straightforward tasks leave little room for behavioural diversity. For more complex tasks and across all n-gram levels, \agent consistently exhibits higher entropy. \revision{To validate this statistically, we conducted one-tailed paired Student’s t-tests for each task (Table~\ref{tab:MC_t_test}). All p-values are below 0.05, except for the simple tasks, confirming that \agent achieves significantly greater solution diversity than ODYSSEY in complex tasks.} Overall, these results show that \agent generates diverse action sequences for complex scenarios while maintaining high success rates.

\begin{table}[!t]
    \caption{\revision{
    \textbf{Results of the one-tailed paired Student’s t-test.} 
    The null hypothesis is that \agentIn{'s} solution diversity is less than or equal to ODYSSEY’s for action subsequences of length $n \in [1,3]$. 
    P-values $< 0.05$ (marked with “*”) reject the null, showing that \agent achieves higher solution diversity.
}}
    \label{tab:MC_t_test}
    \centering
    \revision{\begin{tabular}{lc}
        \toprule
        \textbf{Task} & \textbf{P-value} \\
        \midrule
        Make 1 Sugar & 0.643 \\
        Shear 1 Sheep & 0.939 \\
        Cook 1 Meat & 0.0132* \\
        Obtain 1 Diamond & 0.0487* \\
        Combat 1 Cave Spider & 0.0220* \\
        Combat 1 Skeleton & 0.0170* \\
        Combat 1 Spider & 0.0166* \\
        Survive 1 Day & 0.0247* \\
        \bottomrule
    \end{tabular}}

\end{table}

\subsubsection{\textbf{Discussion}}\label{subsec:comparison-d}

Unlike \agentIn{,} which dynamically generates code, ODYSSEY directly invokes pre-coded functions (Skills) that align with the current plan. While this allows ODYSSEY to execute actions instantly, \agent requires more time to generate and refine code during execution.

Despite its faster performance, ODYSSEY demands a significantly larger upfront investment to develop and maintain a comprehensive Skill Library. This dependency also limits its flexibility in handling unforeseen or novel tasks generated by the Planner, reducing its capacity to test diverse interactions. In several cases, this led to imprecise task execution. For instance, when the plan specified ``smelt iron ore into iron ingots'' but no matching function was available, ODYSSEY repeatedly invoked the unrelated ``mine raw iron'' function based on similarity scores, resulting in incorrect and incomplete execution.
This limitation also explains ODYSSEY's higher solution diversity in two simpler tasks, where it repeatedly performs unrelated actions due to misaligned function invocation.
To confirm this limitation of ODYSSEY, we replaced the Skill Library of ODYSSEY with our example Skills. The results show that it failed to solve any tasks beyond the specified Skills. This constraint ultimately reduces its flexibility and precision in problem-solving and game-testing scenarios.

\section{Related Work}\label{sec:related-work}

\subsubsection*{\textbf{Machine-Learning-Based Game Agents}}
Recent advances in game agents have leveraged reinforcement learning (RL) and imitation learning (IL). RL agents like OpenAI Five~\cite{OpenAI5DefeatsDota2Team} and AlphaStar~\cite{AlphaStar2023} excel in real-time strategy games via self-learning, often surpassing human players. IL approaches like AlphaGo~\cite{AlphaGo2017} and VPT~\cite{48VPT2022} improve learning efficiency by mimicking expert gameplay.
However, both methods rely on primitive actions and predefined rewards or demonstrations, restricting exploration and adaptability~\cite{Go-Explore2021-11, EvolvingMultimodal2019-12}, making such tools often fail to explore diverse behaviours in games systematically.
Furthermore, RL tools often depend on developer-defined rewards~\cite{ExplorationRL2022}, introducing expert bias, whereas IL tools rely heavily on human demonstrations, limiting flexibility. Both methods often operate as black boxes with little interpretability of the decision-making process~\cite{61AutoMC2024}, hindering generalization to novel tasks.

Our approach addresses these gaps by integrating personality into an LLM-based Hybrid Planner that supports diverse strategies and richer exploration. Unlike prior black-box models, \agent offers transparent reasoning via prompt chaining, and demonstrates strong cross-game and cross-scenario adaptability through successful deployments and effective performance in three different games.

\subsubsection*{\textbf{LLM-Based Game Agents}}

Recent advances highlight the potential of LLMs in game agents. Frameworks like ReAct~\cite{77ReAct2023} showcase LLMs' planning ability under dynamic conditions, while DEPS~\cite{81DEPS}, ODYSSEY~\cite{56Odyssey2024}, Voyager~\cite{Voyager2023}, and others~\cite{49Jarvis-1_2023, 52Steve-1_2024, GITM2023} tackle complex tasks in a large-scale game, Minecraft, building on prior successes in domains from board games~\cite {ChessGPT2023} to video games~\cite {75LLMStarCraft}. Hybrid approaches like Auto MC-Reward~\cite{61AutoMC2024}, combining RL with LLM planning, further enhance adaptability and performance.

Most prior work optimizes agents for task completion, often producing homogeneous and repetitive behaviours. In contrast, we introduce \agentIn{,} an LLM-based agent explicitly designed for game testing. Leveraging a personality-driven Hybrid Planner, \agent adapts to varied environments through distinct personalities, promoting diverse exploration and mirroring how human testers evaluate games.

\subsubsection*{\textbf{Agents Mimicking Human Behaviour}}
Previous studies have shown that LLMs can mimic personality traits~\cite{78LLMPotentialOnPersonality2024, 79PersonalityTraitsLLM2023, 80LLMSimulateBigFivePersonality2024, 2022WhoIsGPT3}, primarily through question-answering tasks. While effective in static text-based settings, these approaches lack functional agent implementations or practical applications. In contrast, \agent extends personality mimicking into games, enabling actionable behaviours beyond question answering.

Google’s 2023 study~\cite{1GAVillage2023} explored multi-agent simulations using LLMs with memory and social reasoning in a sandbox world. While showcasing complex social behaviour, it was limited to sandbox simulations without adaptability to real games. \agent builds on these ideas with a deployable, game-agnostic framework for testing real-world games through memory-driven, personality-conditioned planning.

PathOS~\cite{ArtificialPlayers2020}, an RL-based agent, modelled personality through reward shaping for level design but focused only on navigation and lacked a full planning–execution pipeline. It also suffered typical RL limitations: restricted exploration, low interpretability, and poor generalization from manually tuned rewards. \agent addresses these gaps by combining LLM-based planning with a Memory System and Summarizer, enabling transparent, adaptive decision-making and flexible generalization across diverse games.

\section{Threats To Validity} \label{threats-to-validity}

\subsubsection*{\textbf{LLM Selection}} 
The choice of LLM may impact the validity of \agentIn{'s} performance. This study uses GPT-4o (version 2024-08-06)~\cite{OpenAIGPT4o} via API, incurring latency and monetary cost (thousands of USD for all experiments). Training data also limited its precision in game-specific reasoning and personality mimicking. Future work will address these challenges with locally hosted, fine-tuned LLMs optimized for gameplay and personality alignment. Despite these constraints, \agent achieved superior results in automated game testing.

\subsubsection*{\textbf{Personality Trait Definitions}}

The personality prompts used by the Planner may influence the validity of \agentIn{'s} performance. The seven personalities in \agent are directly taken from PathOS~\cite{ArtificialPlayers2020}, which synthesizes traits grounded in real player behaviours from prior research. \revision{These traits and the behaviours of \agent were not independently validated against real player data, as \agentIn{'s} goal is not to reproduce ground-truth human behaviour but to integrate personality traits to generate diverse behaviours. Although the prompts may not perfectly capture each trait, and LLMs cannot fully replicate human contextual adaptability, our evaluation shows that incorporating personality traits substantially increases action diversity. In practice, \agent consistently exhibited behaviours that were distinct and aligned with their respective traits across similar scenarios, demonstrating \agentIn{'s} effectiveness despite this limitation.}

\subsubsection*{\textbf{Game Subject Selection}}

The selection of game subjects may affect the generalizability of our findings. This study focused on games that vary in scale and type but are limited to non-time-sensitive RPGs, specifically, dungeon crawlers, one of the most popular subgenres in this category. Future work will expand to other game types to further validate \agentIn{’s} adaptability across broader gameplay contexts.

\subsubsection*{\textbf{\revision{Human Tester Sample Size}}}
\revision{
    We recruited seven human testers for the evaluation group. While a larger pool would improve generalizability, our sample size was limited by budget constraints.
    In DA, testers show identical code/branch coverage with variation in combinatorial coverage (Table~\ref{tab:human-baseline-variance}). Because DA has few elements to cover, the numerical differences between testers remain small, making the influence on generalizability marginal. In SPD, variance appears across all coverage types. Although absolute variance in combinatorial coverage is small, the low overall human coverage makes relative variation appear larger.
    Since human testers serve only as a baseline for assessing \agentIn{,} the variance is not the focus of our evaluation.
    Nevertheless, when combined with the confidence interval comparisons in Section~\ref{sec:effective-evaluation}, the substantial gap between \agent and humans in SPD remains robust. 
    This suggests that additional human runs are unlikely to change the conclusion that, in complex environments, significant performance gaps persist between them.
}

\subsubsection*{\textbf{Bug Detection Limitations}}
While \agentIn{'s} Summarizer iteratively analyzes outputs during interaction, it can fail to detect or handle in-game bugs effectively, causing repeated task failures without crashing the game. Future work will incorporate game state analysis to improve bug detection. Meanwhile, developers should review consistently failed plans by \agent to identify potential bugs.

\begin{table}[!t]
    \caption{\textbf{\revision{Standard Deviation and Variance Across Seven Human Runs.}} 
    }
    \label{tab:human-baseline-variance}
    \centering
    
    \revision{\begin{tabular}{lccc}
    \toprule
        \textbf{Game} & \textbf{Coverage Type} & \textbf{Standard Deviation} & \textbf{Variance} \\ 
        \midrule
        \multirow{3}{*}{\textbf{DA}} 
        & Code & 0.00\% & 0.00\% \\ 
        & Branch & 0.00\% & 0.00\% \\ 
        & Combinatorial & 4.45\% & 19.76\% \\  \midrule

        \multirow{3}{*}{\textbf{SPD}} 
        & Code & 3.70\% & 13.71\% \\
        & Branch & 3.14\% & 9.85\% \\
        & Combinatorial & 0.093\% & 0.87\% \\
        
    \bottomrule
    \end{tabular}}
\end{table}

\section{\revision{Future Work \& Implications}}

\revision{Our findings highlight both the promise and the limitations of personality-driven agents for automated game testing. While \agent demonstrates clear improvements in behavioural diversity and coverage, humans still outperform it across many dimensions, revealing a substantial gap. Bridging this gap offers opportunities to advance game testing toward better reflecting diverse user behaviours.}
    
\revision{A natural next step is to move beyond fixed personality prompts toward adaptive profiles that evolve with context. Leveraging its flexibility in integrating diverse personality forms, \agent can adapt in multiple ways. For example, the Memory System could periodically reflect on experiences to adjust behaviours over time, while traits could be learned from real player trajectories, with fine-tuned LLMs supporting more authentic evolution. These directions bring agents closer to human-like adaptability and position \agent as a foundation for broader research on personality-aligned game agents.}

\revision{
    Another key direction is to broaden the environments in which such agents operate. Deploying \agent in a new game involves only minor adaptations to the Planner and the Action Executor. For the Planner, the prompt is adjusted to describe the game’s mechanics and map in-game elements to personality traits (Section~\ref{subsubsec:approach-gamer-personality-mimicking}). For the Executor, the available APIs should be exposed to allow \agentIn{’s} interaction with the environment. We are also deploying \agent to additional game types, including larger-scale games such as non-turn-based RPGs and Massively Multiplayer Online (MMO) games.
    
    However, inference latency limits its applicability in time-sensitive game types (e.g., First Person Shooter (FPS)). In our experiments, each action averaged 12.4 seconds, making \agent impractical for real-time deployment. The monetary cost is at \$0.06 USD/\$0.05 USD with/without code generation. At this rate, gameplay sessions with thousands of actions (e.g., SPD) could become expensive, limiting \agent for larger-scale use. To mitigate these challenges, we are exploring smaller, fine-tuned local models, and as LLMs continue to become faster and cheaper, MIMIC can play an increasingly impactful role in automated game testing.
}

\revision{\agentIn{’s} results show that integrating personality traits into automated agents substantially improves behavioural diversity and coverage in game testing. For practitioners, \agent provides a practical tool to uncover edge cases and diverse usage patterns. For researchers, it introduces a new methodology for designing and evaluating test agents with personality in mind.

Looking forward, we see opportunities to expand this line of work. Beyond the direct directions already discussed, one avenue is to extend personality-driven agents within games, enriching user experiences through more realistic Non-Player Characters (NPCs). Beyond games, personality-aware automation also applies to domains like User Interface (UI) testing and Human-Computer Interactions (HCIs), where diverse navigation paths help uncover edge behaviours that traditional tools may miss. In this sense, \agent marks a step toward behaviourally rich, personality-aware testing methodologies across automated software engineering practice.}

\section{Conclusion}\label{sec:conclusion}

Inspired by the diverse strategies of human players during gameplay, we introduced \agentIn{,} a novel testing framework that integrates personality traits into LLM-based gaming agents. Utilizing a Hybrid Planner to emulate varied in-game behaviours through a Memory System that accumulates experience, \agent enables personality-aligned decision-making, enhancing behavioural diversity and testing effectiveness.

We validated \agent on two open-source games of varying complexity, where it consistently outperformed random-based baselines and ablated versions in both code and interaction-level coverage. In Minecraft, it also surpassed a state-of-the-art LLM agent, achieving higher task success and greater strategic diversity. These results highlight \agentIn{’s} ability to generate personality-driven actions across environments and drive the exploration toward more diverse scenarios.

Overall, \agent advances automated game testing with a scalable, personality-driven framework that adapts to dynamic environments and goals. Our results confirm its effectiveness and establish \agent as a powerful, generalizable tool for testing modern, complex games at scale.

Furthermore, \agent offers direct value by helping practitioners uncover edge cases and diverse usage patterns, and giving researchers a methodology for designing and evaluating personality-driven test agents. In a broader context, this work highlights the potential of personality-aware automation to enrich player experiences in games and to extend its capabilities to other domains such as UI testing and HCI.

\section*{Acknowledgment}

The authors would like to thank ASE 2025 reviewers for their constructive comments.
This work is supported by the Natural Sciences and Engineering Research Council of Canada Discovery Grant (Grant No. RGCPIN-2022-03744 and Grant No. DGECR-2022-00378) and Fonds de recherche du Québec-secteur Nature et technologies (Grant No.363482~\cite{frqntNewAca}).

\newpage

\bibliographystyle{IEEEtran/IEEEtran}
\bibliography{IEEEtran/IEEEabrv,references}

\end{document}